\documentclass[aps,showpacs,preprintnumbers,amsmath,nofootinbib,amssymb]{revtex4}
\usepackage[dvips]{epsfig}
\usepackage{graphicx}
\usepackage{pgf}
\usepackage{hyperref}
\newcommand{\nc}[1]{\newcommand{#1}}

\nc{\its}[1]{\itshape #1 \upshape}
\nc{\mc}[3]{\multicolumn{#1}{#2}{#3}}

\nc{\bc}{\begin{center}}
\nc{\ec}{\end{center}}
\nc{\ig}[1]{\bc \includegraphics{#1} \ec}

\nc{\bo}[1]{\mbox{\boldmath \( #1 \! \! \)  \unboldmath}}
\nc{\be}{\begin{eqnarray}}
\nc{\ee}{\end{eqnarray}}
\nc{\bew}{\begin{eqnarray*}}
\nc{\eew}{\end{eqnarray*}}
\nc{\bs}{\begin{subeqnarray}}   
\nc{\es}{\end{subeqnarray}}     
\nc{\nnn}{\nonumber \\}
\nc{\f}[2]{\frac{#1}{#2}}
\nc{\td}[2]{\f{d #1}{d #2}}
\nc{\pd}[2]{\f{\partial #1}{\partial #2}}
\nc{\suli}{\sum\limits}
\nc{\proli}{\prod\limits}
\nc{\ili}{\int\limits}
\nc{\sr}[2]{\stackrel{#1}{#2}}
\nc{\dps}{\displaystyle}
\nc{\ket}[1]{\left| #1 \right>}
\nc{\bra}[1]{\left< #1 \right|}
\nc{\bracket}[2]{\left< #1 \right| \left. \! #2 \right>}
\nc{\norm}[1]{\left\| #1 \right\|}
\nc{\lndm}[1]{\pd{^{#1} \ln{\det{M}}}{\mu^{#1}}}
\nc{\pdmm}[1]{M^{-1} \pd{^{#1} M}{\mu^{#1}}}
\nc{\pdm}{M^{-1}\pd{M}{\mu}}
\nc{\trac}[1]{\mbox{Tr}\left(#1\right)}
\nc{\hm}{\hat{m}}

\def\lsim{\raise0.3ex\hbox{$<$\kern-0.75em\raise-1.1ex\hbox{$\sim$}}}
\def\gsim{\raise0.3ex\hbox{$>$\kern-0.75em\raise-1.1ex\hbox{$\sim$}}}

\begin{document}

\title{Chiral phase structure of three flavor QCD at vanishing baryon number density}

\author{
A. Bazavov$^{\rm a}$, H.-T. Ding$^{\rm b}$, P. Hegde$^{\rm c}$,
F. Karsch$^{\rm d,e}$, E. Laermann$^{\rm d}$, \\Swagato Mukherjee$^{\rm e}$, P. Petreczky$^{\rm e}$ and C. Schmidt$^{\rm d}$
}

\affiliation{
$^{\rm a}$Department of Computational Mathematics, Science and Engineering,
Department of Physics and Astronomy,
Michigan State University, East Lansing, Michigan, 48824, USA\\
$^{\rm b}$Key Laboratory of Quark \& Lepton Physics (MOE) and Institute of
Particle Physics, \\
Central China Normal University, Wuhan 430079, China\\
$^{\rm c}$Center for High Energy Physics, Indian Institute of Science, Bangalore 560012, India \\
$^{\rm d}$Fakult\"at f\"ur Physik, Universit\"at Bielefeld, D-33615 Bielefeld,
Germany \\
$^{\rm e}$Physics Department, Brookhaven National Laboratory, 
Upton, NY 11973, USA
}

\date{\today}

\begin{abstract}
We investigate the phase structure of QCD with 3 degenerate quark flavors as 
function of the degenerate quark masses at vanishing baryon number density.
We use the
Highly Improved Staggered Quarks on lattices with temporal extent $N_{t}=6$
and perform calculations for six values of quark masses,  
which in the continuum limit correspond to pion masses in the range
$80~{\rm MeV} \lesssim m_{\pi} \lesssim 230~$MeV. 
By analyzing the volume and temperature dependence of the chiral condensate
and chiral susceptibility we find no direct evidence for a first order phase 
transition in this range of pion mass values. 
 Relying on the universal scaling behaviors of the chiral observables near an anticipated chiral critical point, 
 we estimate an upper bound for the critical pion mass, $m_\pi^c \lesssim$ 50 MeV,  below which a region 
 of first order chiral phase transition is favored.

\end{abstract}


\maketitle

\section{Introduction}

Mapping out the QCD phase diagram is one of the basic goals of lattice
QCD calculations at non-zero temperature. It was noted by Pisarski and 
Wilczek that the order of the chiral phase transition in QCD may depend on the 
number of light quark degrees of freedom and qualitative features of 
the transition
may also change with the quark mass \cite{Pisarski:1983ms}. 
In QCD with 3 massless quark flavors the chiral phase transition is expected to
be first order. If this is the case, the phase transition remains first order 
even for non-zero values of the
quark masses and terminates at a critical quark mass 
$m_q^c$, or equivalently at a critical pion mass, $m_\pi^c$, 
where the transition becomes second order belonging to the 3-d Z(2) Ising universality class. For quark mass $m_q > m_q^c$  
chiral restoration takes place through a smooth crossover.

Knowledge about the phase structure in the light-strange quark mass plane at vanishing 
chemical potential also impacts our understanding regarding its extension to non-zero chemical potential. 
For zero chemical potential, the values of critical quark masses characterize a line of chiral phase transitions 
in the light-strange quark mass plane. This line extends toward the non-zero chemical potential direction and forms 
a surface of phase transitions in the 3-d, Z(2) universality class. The chemical potential where this surface intersects the physical values of 
light and strange quark masses may correspond to the QCD critical point \cite{Schmidt:2002uk}. However,
determining the curvature of this surface turns out to be complicated
\cite{deForcrand:2008vr,Jin:2015taa} and, in fact, is likely to suffer
from similar lattice cut-off effects as those contributing to the 
value of the critical pion mass itself\footnote{Other possibilities for 
generating a second order transition at the physical values
of quark masses have been discussed in Ref.~\cite{Gupta:2008ac}.}.

The first order chiral phase transition in 3--flavor QCD has been investigated on 
coarse lattices using unimproved~\cite{Christ:2003jk, deForcrand:2003ut,deForcrand:2007rq,Smith:2011pm} as well as improved actions
~\cite{Karsch:2001nf,Schmidt:2002uk,Karsch:2003va, Bernard:2004je,Cheng:2006aj,Jin:2014hea,Takeda:2016vfj}.
However, on these coarse lattices the critical pion mass value turns out to
be strongly cut-off and regularization scheme  dependent.  
As of now no continuum extrapolated results
exist. Current results for the critical pion mass obtained in calculations with staggered (standard and p4fat3)
fermions on $N_\tau=4$ and 6 lattices vary from about 300 MeV down to about 70 MeV~\cite{Christ:2003jk, deForcrand:2003ut,deForcrand:2007rq,Smith:2011pm,Karsch:2001nf,Schmidt:2002uk,Karsch:2003va, Bernard:2004je,Cheng:2006aj}. While studies using the clover improved
Wilson fermion action on $N_t=4$, 6, 8 and 10 lattices suggest that $m_\pi^c$ can change from about 750 MeV to about 100 MeV~\cite{Jin:2014hea,Takeda:2016vfj}. In general it is found that the critical pion mass decreases 
when using either improved actions or when reducing the lattice
spacing\footnote{
A study of 4-flavor QCD using HYP action~\cite{Hasenfratz:2001ef} also suggests that the first order chiral phase transition becomes weaker in the continuum limit.}.

In addition to studies concerning $N_f$=3,  lattice QCD calculations searching for first order chiral phase transitions in other cases have also been carried out.
Studies on $N_f$=2 QCD using standard staggered fermions~\cite{Bonati:2014kpa} and unimproved Wilson fermions~\cite{Philipsen:2016hkv} on $N_t=4$ lattices
suggest that $m_\pi^c$ is nonzero and could be around 560 MeV. Investigations on $N_{t}=6$ lattices have
also been performed using an improved staggered fermion action (2stout action), 
in which the three quark flavors are not taken to be degenerate. Instead the ratio 
of light to strange quark masses has been kept fixed to about 1/27 when approaching 
the massless limit~\cite{Endrodi:2007gc}.  The analysis of Ref.~\cite{Endrodi:2007gc} suggests that $m_\pi^c$ is around $50 $ MeV. 
 While a recent study on QCD with two degenerate light quarks approaching to chiral limit 
 and a fixed physical strange quark mass using the 
Highly Improved Staggered Quark (HISQ) action
 on $N_\tau=6$ lattices suggests
 that $m_\pi^c$ is compatible with zero~\cite{Ding:2015pmg}.

To advance the understanding of the chiral phase transition in 
3-flavor QCD we study the chiral phase structure at vanishing baryon density 
using the HISQ action on lattices
with temporal extent $N_\tau=6$.
Preliminary results have been reported in conference proceedings~\cite{Ding:2011du,Ding:2013nha,Ding:2015pmg}. 
The paper is organized as follows. In section~\ref{sec:lattice} we give 
details on the parameters used in our calculations.
In section~\ref{sec:universality} we describe the universal properties in the vicinity
of the chiral phase transition and
chiral observables. In section~\ref{sec:results} we present our results on the 
phase structure of 3-flavor QCD
and finally we summarize in section~\ref{sec:summary}.

\section{Lattice formulation and setup}
\label{sec:lattice}

It has been noted previously that the estimate of the critical pion mass $m_\pi^c$ from lattice QCD
calculations strongly
depends on lattice cutoff effects. In order to reach a better understanding of 
the first order chiral phase transition region 
we use here the HISQ action~\cite{Follana:2006rc}.
At a given value of the lattice spacing the HISQ action
achieves better taste symmetry than the asqtad, p4 and 2stout actions, which previously
have been used for the analysis of the phase structure of 3-flavor QCD~\cite{Ding:2015ona}. 
Furthermore, we use a
tree-level improved Symanzik gauge action and perform 
calculations on lattices with temporal extent $N_t$=6. 
The calculations have been performed for 3 degenerate quark flavors at
six different values of quark masses $m_q$ (in units of the lattice spacing) in the range
$0.0009375 \le m_q\le 0.0075$. 
Gauge configurations have been generated with a rational hybrid Monte
Carlo (RHMC) algorithm.
The chiral observables (introduced below) were measured after every 10th trajectory of unit length using 40 Gaussian-distributed stochastic sources.

To convert our simulation parameters to physical units
we use results on the determination of the lattice spacings in
(2+1)-flavor QCD obtained by the HotQCD
collaboration  \cite{Bazavov:2011nk,Bazavov:2014pvz}.
The lattice spacing is fixed using the $r_0$ scale~\cite{Bazavov:2011nk}  
and hadron masses 
calculated on lines of constant physics. To be specific, for a fixed 
ratio of the light to strange quark masses, $m_l/m_s=1/20$, 
the hadron masses have 
been determined by demanding that the strange quark mass attains its physical value.
In order to use this for our 3-flavor analysis we take into account that
the lattice spacing at a given value of the gauge coupling depends on the
quark masses, i.e. as we vary the quark masses the value of $r_0$ in 
lattice units, $r_0/a$, will change. Using results for $r_0/a$ obtained
in calculations for (2+1)-flavor QCD with two different values of the
light to strange quark mass ratio, $m_l=m_s/5$ \cite{Bazavov:2011nk}
and $m_l=m_s/20$ \cite{Bazavov:2011nk,Bazavov:2014pvz},
we can estimate the dependence of $r_0/a$ at fixed values of the 
gauge coupling $\beta=10/g^2$, on the quark mass combination $m_s+ 2 m_l$. 
Assuming that  
$r_0/a$ only depends on this combination of the quark masses we can 
estimate its value for the case of three degenerate flavors for different 
values of the quark mass, $m_q$. 
 We find that for the lightest quark mass $m_q = m_s /80=0.0009375$ 
 the change in $r_0/a$ in the relevant range of gauge couplings $5.8 \leq \beta \leq 6.1$ 
 amounts to $\sim 4\%$. For example, for $N_f=3$ with $m_q=0.0009375$ and close to 
 the critical coupling $\beta=5.85$ the estimated value is $r_0/a=1.82$. In comparison, the corresponding 
 value for $N_f = 2 + 1$ with physical $m_s$ and $m_l=m_s/20$ is $r_0/a=1.75$.

Since the dependence of the lattice scale on quark masses seems to be 
moderate down to the smallest quark mass 
value used in our study, we estimate the value of pseudo-scalar meson masses 
relevant for our choice of parameters by simply rescaling the pion mass 
obtained in the (2+1)-flavor studies by the corresponding
ratio of the light quark masses, i.e. by $\sqrt{m_q/m_l}$.
From this we find that the range of quark mass values explored by
us corresponds to masses for the lightest pseudo-scalar meson in the range
$80~{\rm MeV} \lesssim m_{\pi} \lesssim 230~$MeV. In the continuum limit
this corresponds to the value of the pion mass.
For the spatial size of the lattice we generally use $N_s=24$. This ensures 
that the product of pion mass and spatial extent $L=N_s a$ stays large also
for the lightest quark masses, i.e. $m_\pi L\gtrsim$ 3. 
At the second largest and smallest value of the quark masses used by us,
$m_q=0.00375$ and $m_q=0.0009375$, respectively, we also performed simulations 
for different spatial lattice sizes. We used
four different lattice sizes for the second heaviest quark mass, $N_s=10$, 12, 16 and 24, and two lattice sizes for the lightest quark mass, $N_s=16$ and 24.
The simulation parameters and corresponding pion masses in the continuum limit are listed in Table~\ref{tab:parameter}.

\begin{table}[th]
\begin{center}
\vspace{0.3cm}
\begin{tabular}{|c|c|c|c|c|}
\hline
$N_s^3\times N_t$ &   $m_q$     & $m_{\pi}$ [MeV]     &\# $\beta$ values & average \# of conf. for each $\beta$\\
\hline
$16^3\times$ 6           & 0.0075          & 230                       &  9                    & 1000             \\
$24^3\times$ 6          & 0.00375        & 160                    &  11                  & 2000             \\
$16^3\times$ 6          & 0.00375        & 160                    &  8                    & 2000             \\
$12^3\times$ 6          & 0.00375        & 160                    &  9                    & 2000             \\
$10^3\times$ 6          & 0.00375        & 160                    &  11                    & 1500             \\
$24^3\times$ 6          & 0.0025        &  130                       &   7                     & 1300          \\
$24^3\times$ 6           & 0.001875      & 110                      &  8                    & 1000           \\
$24^3\times$ 6          & 0.00125        &  90                       &   7                     & 1000           \\
$24^3\times$ 6           & 0.0009375    & 80                       &   8                     & 1500        \\
$16^3\times$ 6        & 0.0009375    &  80                   &   8                   & 1500          \\
\hline
\end{tabular}
\end{center}
\caption{
Parameters used in simulations of 3-flavor QCD using the HISQ action and the 
tree-level improved Symanzik gauge action. 
}
\label{tab:parameter}
\end{table}

The basic observables used in our analysis are the chiral condensate 

\be
\frac{\langle \bar{\psi}\psi \rangle_{\rm q}}{T^3} &=& \frac{1}{3} \frac{1}{VT^2}
\frac{\partial \ln Z}{\partial m_q}
= \frac{N_t^2}{4 N_s^{3}} 
\left\langle {\rm Tr}\; D^{-1}(m_q)\right\rangle,
\label{orderparameter1}  
\ee
and the disconnected part of the chiral susceptibility
\be
\frac{\chi_{\rm q,disc}}{T^2} &\equiv& \frac{N_t}{16N_s^{3}} \left( 
\left\langle \left( {\rm Tr}\; D^{-1}(m_q)\right)^2 \right\rangle -
\left\langle {\rm Tr\;} D^{-1}(m_q)\right\rangle^2\right) 
, \label{sus_chi}
\ee
where $Z$ denotes the QCD partition function and $D$ is the staggered fermion 
matrix. 
The chiral condensate and the chiral susceptibility are
normalized to one flavor degree of freedom.

\section{Universal properties near a critical point}
\label{sec:universality}

In the vicinity of a critical point the free energy of a system can be 
expressed as a sum of a singular and a regular part,

\begin{equation}
f = - \frac{T}{V}\ln Z \equiv f_{\rm sing}(T,m_q) + f_{\rm reg}(T,m_q).
\label{free}
\end{equation}
The singular contribution is given in terms of a scaling function and 
critical exponents characteristic for the universality class of the critical point,
\begin{equation}
f_{\rm sing}(T,m_q) = h_0 h^{1+1/\delta} f_s(z)\; ,\; z= t/h^{1/\beta\delta} \; .
\label{sing}
\end{equation}
Here $\beta$ and $\delta$ are universal critical exponents, $h_0$ is a non-universal 
normalization factor,  $t$ and $h$ are reduced 
temperature and symmetry breaking parameters, respectively. They vanish at 
the critical point, $(t,h)=(0,0)$, and are functions of the couplings, 
$T$ and $m$. For the 3-dimensional Z(2)
universality class the critical exponents $\beta=0.3207$, $\delta=4.7898$
and $\gamma= \beta(\delta -1)=1.2371$~\cite{El-Showk:2014dwa}.

The singular part of the free energy density, $f_{\rm sing}(T,m)$, dominates over 
the regular part 
when the system is close to the critical region. The order parameter $M$ 
of the transition and its susceptibility $\chi_M$ are then governed by 
scaling functions that arise from the scaling form of the singular part 
of the free energy~\cite{Ejiri:2009ac,Karsch:2010ya}
\begin{eqnarray}
M(t,h)&=&-\partial f_{\rm sing}(t,h)/\partial H = h^{1/\delta}\, f_{G}(z),
\label{eq:scaling} \\
\chi_M &=& \frac{\partial M}{\partial H} = \frac{1}{h_0} h^{1/\delta-1} 
f_\chi(z),
\label{eq:meos_chi}
\end{eqnarray}
where $f_G(z)=-\Big( 1+\frac{1}{\delta}\Big) f_s(z) +\frac{z}{\beta\delta} 
\frac{\partial f_s(z)}{\partial z}$ and 
$f_\chi(z) = \frac{1}{\delta} \Big( f_G(z) - \frac{z}{\beta}\frac{\partial f_G(z)}{\partial z}\Big{)}$
are universal scaling functions.
The variables $t$ and $h$ are related to the temperature 
$T$ and the symmetry breaking (magnetic) field $H\equiv h_0 h$. 
The order parameter susceptibility may be used to introduce a variable $t_p$ as the pseudo-critical
temperature, which  is defined as the location of the maximum of $\chi_M$ obtained
as a function of $t$ for fixed $h$. This is reached for some value
$z_p= t_p/h^{1/\beta\delta}$. One thus finds the standard scaling behavior
of $\chi_M^{peak}$ as a function of the external field $h$
\begin{eqnarray}
\chi_M^{peak} &\sim& h^{1/\delta -1} f_\chi(z_p)\;\; .\label{scalingm} 
\end{eqnarray}

The generic discussion of the scaling behavior of the order parameter 
and its susceptibility given above becomes more complicated in cases
where the scaling variables $t$ and $h$ cannot directly be mapped onto
corresponding couplings of the theory under study, e.g. 3-flavor QCD.
If 3-flavor QCD has a first order transition in the chiral limit,
a second order transition at non-vanishing values
of the quark mass exists, which terminates the line of first
order transitions. This critical endpoint is expected to be in
the $Z(2)$ universality class~\cite{Gavin:1993yk}. 
The relevant fields in the vicinity 
of this critical point can be expressed as linear combinations of 
$m_q-m_q^c$ and $T-T_c$~\cite{Karsch:2000xv}.
Here $T_c$ is the transition temperature at vanishing external 
field $h$, which in 3-flavor QCD calculations is related to a critical 
coupling $\beta_c$ and a critical quark mass $m_q^c$.
Rather than using the temperature $T-T_c$ it is convenient for our discussion  to use the difference of
gauge couplings $\beta=10/g^2$ (not to be confused with the critical
exponent $\beta$)\footnote{In the continuum limit the gauge coupling $\beta$
and the temperature $T=1/(aN_t)$ are related through the asymptotic scaling 
relation, $T/\Lambda = {\rm exp}(\beta/(20b_0))$, with $b_0$ denoting the 
coefficient of the leading term in the QCD $\beta$-function. For small
temperature differences, i.e. in the vicinity of a critical point one thus
finds $(T-T_c)/T_c = \beta-\beta_c$.}. 
The variables $t$ and $h$ may then be related
to the bare couplings of 3-flavor QCD,
\begin{eqnarray}
t&=& (\beta-\beta_c + A\ (m_q-m_q^c))/t_0
\; , \label{eq:mixingt}\\
h&=& (m_q-m_q^c + B\ (\beta-\beta_c) )/h_0 \; .
\label{eq:mixing}
\end{eqnarray}
Although it is not necessarily the case one may assume that the temperature-like
($t$) and external-field-like ($h$) direction are orthogonal to each other.
In that case $B=-A$.

Let us first discuss the scaling behavior of the order parameter
susceptibility in terms of the bare QCD parameters $\Delta m\equiv  m_q-m_q^c$
and $\Delta \beta \equiv \beta-\beta_c$. 
In the $(\Delta \beta )$-$(\Delta m)$ coordinate frame  the constant value
of the scaling variable $z$ is given by
\begin{equation}
z_p = z_0\,\frac{\Delta \beta -B \Delta m}{\left( \Delta m +B \Delta \beta\right)^{1/\beta\delta}} \;\; .
\label{eq:zp}
\end{equation}
Here $z_0=h_0^{1/\beta\delta}/t_0$. The above equation fixes the relation between $\Delta \beta$ and $\Delta m$ required
to keep $z_p$ constant.
Obviously for $B=0$ one just recovers the scaling relation
$\Delta m = (z_0/z_p \,\Delta \beta)^{\beta\delta}$. For $B\ne 0$
we obtain for $\Delta \beta \rightarrow 0$, 
\begin{equation}
\Delta m = -B \Delta \beta + \left( \frac{z_0}{z_p} (1+B^2) 
\Delta \beta \right)^{\beta\delta} +{\cal O} ((\Delta \beta)^{2\beta\delta-1})\; .
\label{eq:mb}
\end{equation}
As $\beta\delta > 1$ for the universality classes of interest to us
the first term in this relation will always
dominate in the limit $\Delta \beta \rightarrow 0$ and one finds
from Eqs.~(\ref{scalingm})-(\ref{eq:mixing}),
\begin{equation}
\chi_M^{peak}\sim h^{1/\delta-1} \sim
\begin{cases}
(\Delta m)^{-(1-1/\delta)} &,\;\; B=0, \cr
(\Delta m)^{-\gamma} &,\;\;  B\ne 0\; .
\end{cases}
\label{eq:chi_singular}
\end{equation}
For any $B\ne 0$ the susceptibility of the order parameter thus will 
diverge with the critical exponent $\gamma$ rather than $1-1/\delta < \gamma$
when approaching the critical point at fixed $z=z_p$. Similarly one finds 
for the order parameter at the critical gauge coupling $\beta_c$,

\begin{equation}
M_c\sim h^{1/\delta} \sim
\begin{cases}
(\Delta m)^{1/\delta} &,\;\; B=0, \cr
(\Delta m)^{\beta} &,\;\;  B\ne 0 \; .
\end{cases}
\label{eq:M_singular}
\end{equation}

In an actual lattice QCD calculation we do not directly deal with the
order parameter $M$ and its susceptibility $\chi_M$. 
The proper order parameter $M$, which detects the breaking of the $Z(2)$
symmetry and vanishes in the symmetry restored phase, can be constructed 
from two independent thermodynamic observables, e.g. a linear combination of the 
chiral condensate $\langle\bar{\psi} \psi \rangle$ and the pure gauge action 
$S_G$
(or second order quark number susceptibility $\chi_2^q$)~\cite{Karsch:2001nf}.
Similarly the susceptibility of the order parameter receives contributions from
several terms, among these is the disconnected part of the chiral 
susceptibility.  Thus it may be expected that the singular behavior of  
chiral condensates $\langle\bar{\psi} \psi \rangle$ and its disconnected chiral
susceptibilities $\chi_q$ obey the relations given in Eqs.~(\ref{eq:M_singular}) and (\ref{eq:chi_singular}), respectively.
In Appendix $A$ we give some 
more details on this and the corrections to scaling that arise from the 
fact that the chiral condensate and its susceptibility are not the correct order parameter 
and order parameter susceptibility for the Z(2) symmetry breaking in 3-flavor QCD.

\vspace{0.2cm}

\section{Results}
\label{sec:results}

\subsection{Chiral condensates and chiral susceptibilities}

\begin{figure}[htp]
\begin{center}
\includegraphics[width=.45\textwidth]{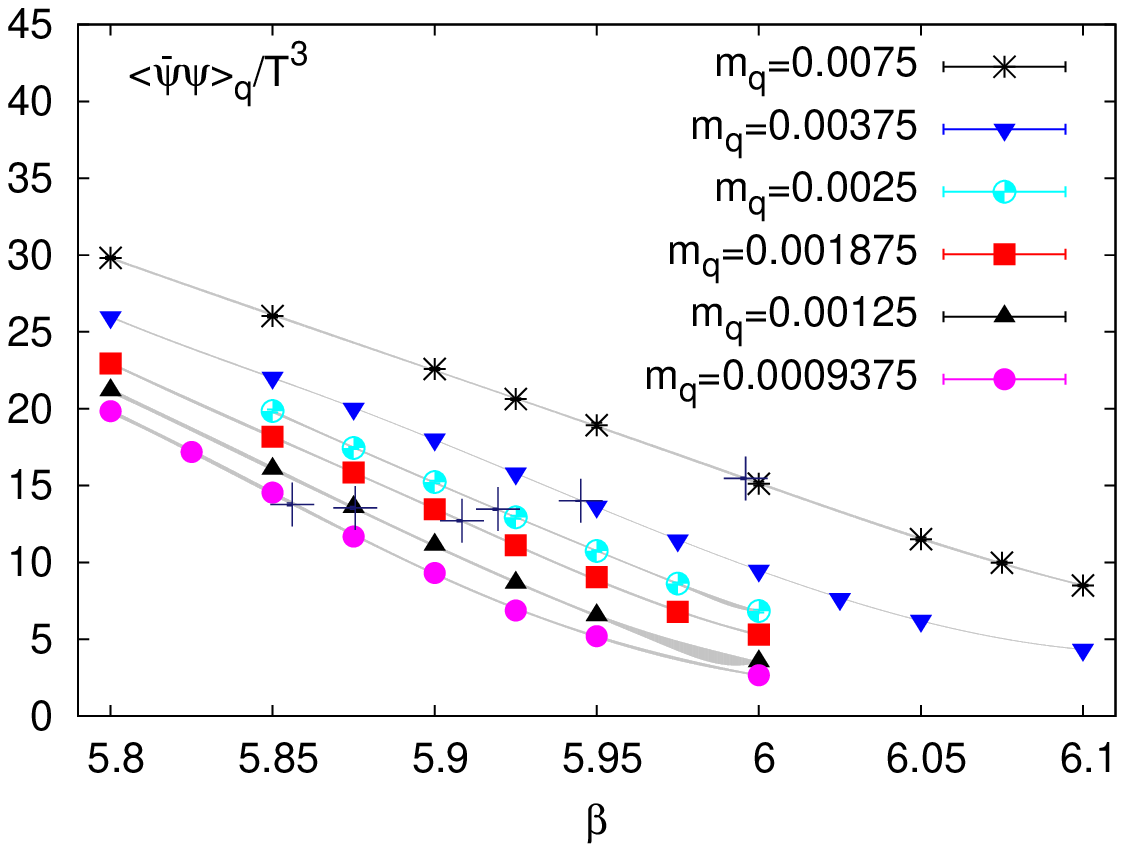}~~\includegraphics[width=.45\textwidth]{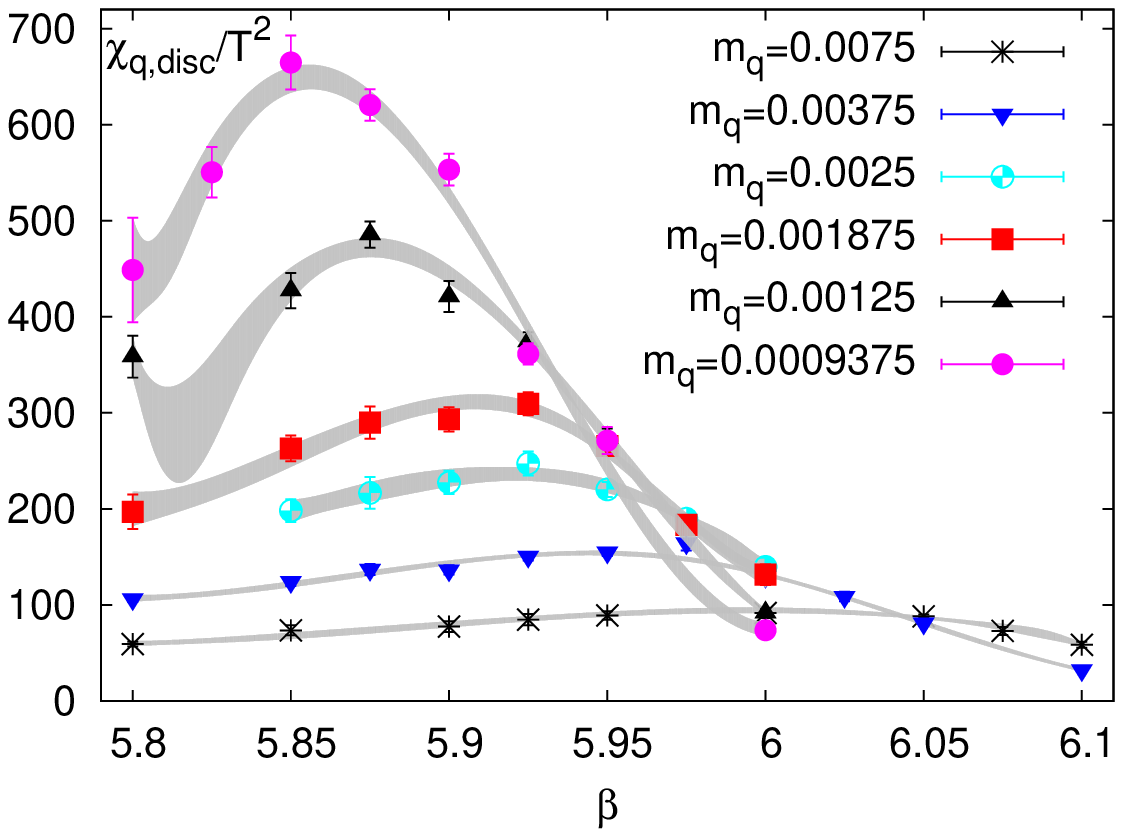}
\end{center}
\caption{The chiral condensates (left) and the disconnected part of the chiral 
susceptibilities (right) versus the gauge coupling $\beta$ for various 
values of the bare quark masses $m_q$.
Shaded curves show spline fits to the chiral observables (for more details
see subsection~\ref{sec:estimate}).  Crosses shown in the left plot indicate 
values for the chiral condensate at the pseudo-critical values of the gauge 
coupling $\beta_c (m_q)$ determined from the location of the peaks of the
disconnected chiral susceptibility.}.
\label{fig:obs_beta}
\end{figure}

In Fig.~\ref{fig:obs_beta}~(left) we show the chiral condensates 
as function of the  gauge coupling $\beta$ 
for various values of the quark masses corresponding to the pion 
masses ranging from $230~$MeV down to $80~$MeV. 
All data shown in this figure
have been obtained on $24^3\times6$ lattices 
except those for the largest quark mass, $m_q=0.0075$, corresponding to  
$m_\pi\simeq 230~$MeV, which are obtained on $16^3\times6$ lattices. 
Obviously, the chiral condensate decreases with increasing value of $\beta$, 
i.e. increasing temperature, 
as well as with decreasing quark mass. 
However, the slope of $\langle\bar{\psi} \psi \rangle$ seems to vary
only little with $\beta$ which differs from the behavior expected from
an order parameter close to a critical temperature. 
This may reflect the fact that $\langle\bar{\psi} \psi \rangle$ is not the true order parameter for the
transition we are trying to probe.

\begin{figure}[htp]
\begin{center}
\includegraphics[width=.45\textwidth]{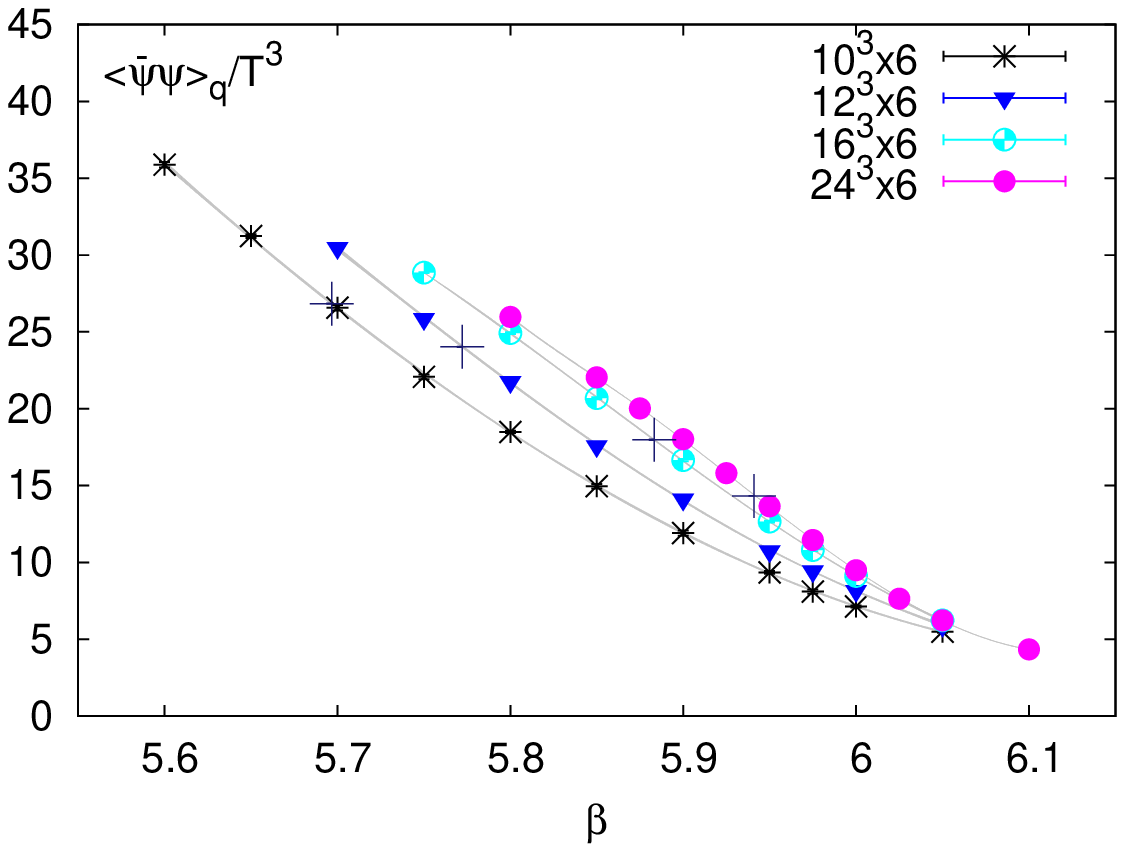}~~~\includegraphics[width=.45\textwidth]{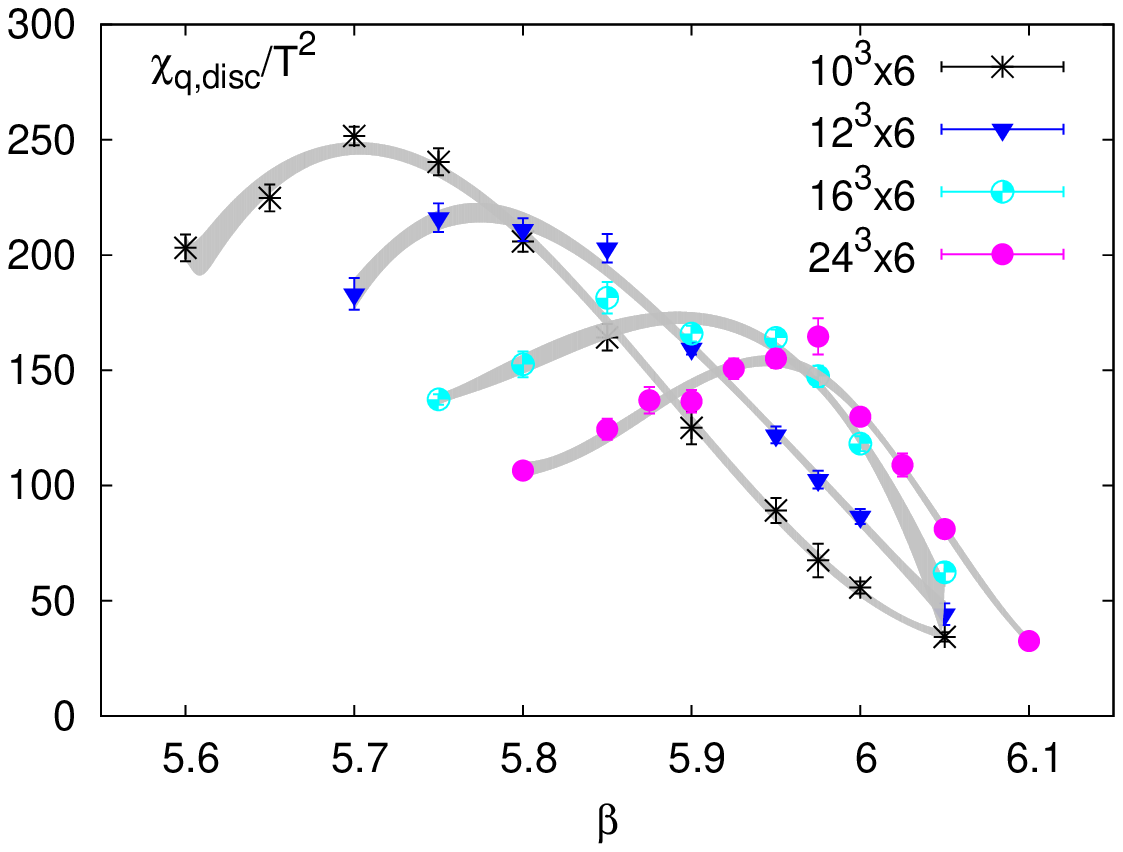}
\end{center}
\caption{Volume dependences of the chiral condensate (left) and the disconnected chiral susceptibility (right) with quark mass $m_q=0.00375$ corresponding to 
$m_\pi\simeq160$ MeV in the continuum limit. Bands in both plots denote interpolations via spline fits and the 'crosses' in the left plot label the critical gauge coupling extracted from the spline fits to
disconnected susceptibilities.}
\label{fig:volume_dep375}
\end{figure}

In Fig.~\ref{fig:obs_beta}~(right) we show the temperature and quark mass dependence of the 
disconnected part of the chiral susceptibilities. They rise with 
decreasing values of the quark mass and has well defined peaks that
shift to smaller values of the coupling $\beta$ as the quark mass decreases,
i.e. the 
pseudo-critical temperature of 3-flavor QCD
decreases with decreasing values of the quark mass. 
The value of the chiral condensate at the pseudo-critical couplings, 
$\beta_c(m_q)$, is also indicated by crosses in Fig.~\ref{fig:obs_beta}~(left).
The decrease of the transition temperature with decreasing value of the 
quark masses is well established in QCD thermodynamics and can 
qualitatively be understood in terms of the quark mass dependence of
hadronic degrees of freedom. With decreasing quark mass they become 
lighter and are thus more easily excited in a thermal heat bath. 
They then contribute to the energy density of the system already at lower
temperatures and can trigger the onset of a phase transition at a 
lower temperature.

The volume dependences of the chiral condensates and chiral susceptibilities at $m_q=0.00375$ are shown in Fig.~\ref{fig:volume_dep375}.
Results obtained for four different volumes, i.e. $N_s$=10, 12, 16 and 24
are presented. As seen 
from Fig.~\ref{fig:volume_dep375}~(left), at a fixed value of the temperature
the chiral condensate increases as the volume is increased and this 
volume dependence is stronger at low temperature than at high 
temperature. The volume dependence of the
disconnected part of the chiral susceptibilities is shown in Fig.~\ref{fig:volume_dep375}~(right). The peak location 
of the disconnected susceptibilities, which defines the pseudo-critical 
temperature, shifts to higher temperatures and the peak height decreases when 
the volume increases. 

\begin{figure}[htp]
\begin{center}
\includegraphics[width=.4\textwidth]{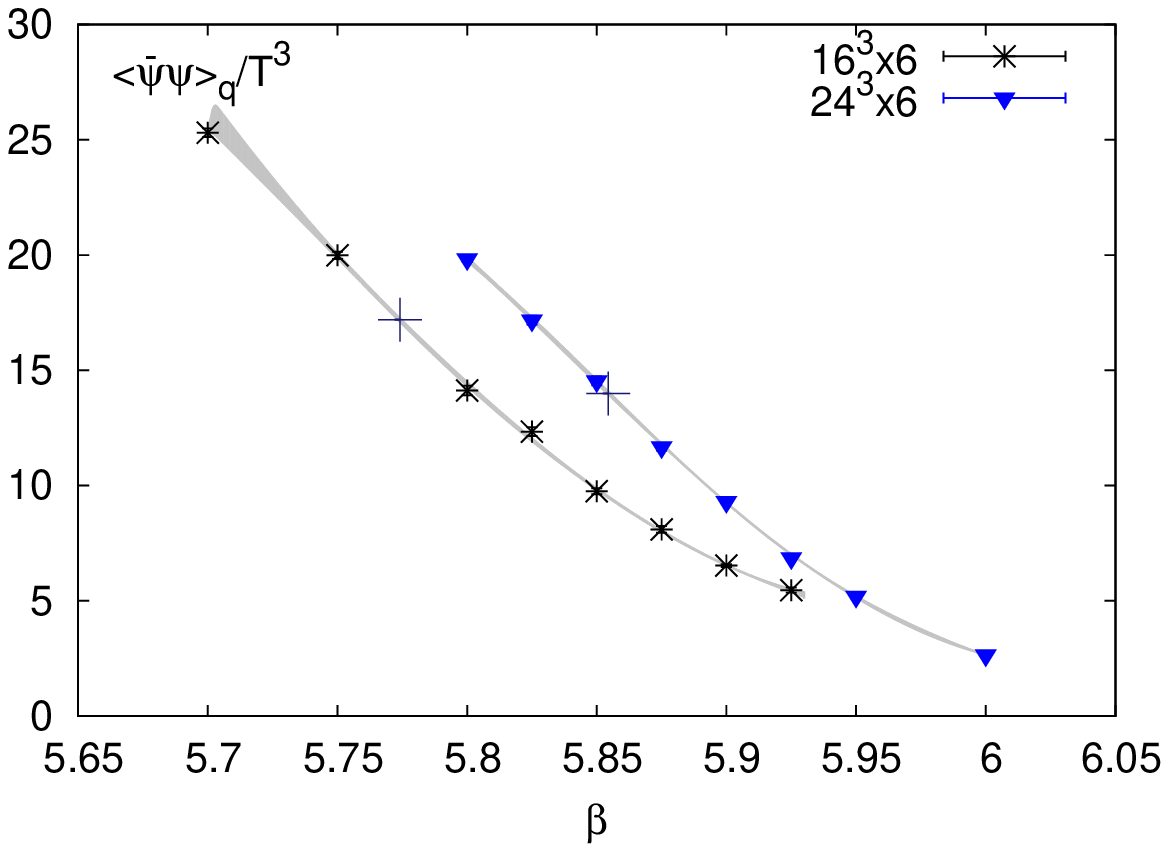}~~~\includegraphics[width=.4\textwidth]{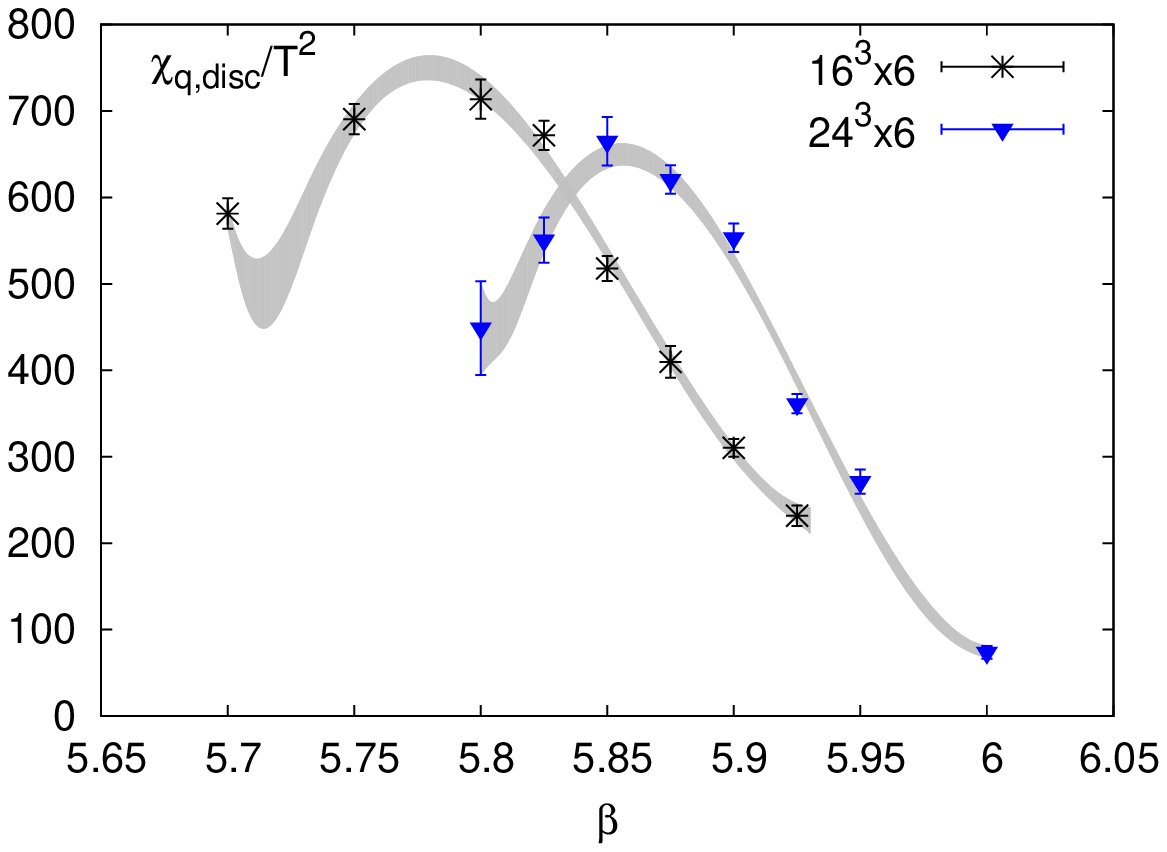}
\end{center}
\caption{Same as Fig.~\ref{fig:volume_dep375} but for $m_q=0.0009375$ corresponding to $m_\pi\simeq80~$MeV in the continuum limit.}
\label{fig:volume_dep9375}
\end{figure}

Corresponding results for the volume dependence of 
$\langle \bar{\psi}\psi\rangle_q$ and $\chi_{q,disc}$ obtained
for the lightest quark mass value, $m_q=0.0009375$
($m_{\pi}\approx 80 $ MeV) are shown in
Fig.~\ref{fig:volume_dep9375}. The pattern seen in the volume dependence of the chiral condensate
and the chiral susceptibility is similar to the one shown for
$m_q=0.00375$ ($m_{\pi}\approx 160 $ MeV) in 
Fig.~\ref{fig:volume_dep375}. In particular, we note that also 
for this small quark mass value the peak height of the chiral
susceptibility decreases with increasing volume.

This volume dependence is consistent with the expected volume dependence 
in the presence of a non-vanishing symmetry breaking field~\cite{Engels:2014bra, Engels:2001bq}. In a
finite volume chiral symmetry is not broken, i.e. the chiral condensate
vanishes in the chiral limit at any value of the temperature. However,
when taking first the infinite volume limit and then the chiral limit
the chiral condensate can approach a non-zero value. One thus expects
that for small values of the quark mass the condensate will increase as 
the volume increases and the volume dependence is larger at low temperatures 
than at high temperatures, because the asymptotic value of the chiral
condensate is larger at low temperatures than at high temperatures.
As the condensate will drop to zero in any finite volume, it also 
varies more rapidly with quark mass, which is reflected by the larger
peak height of the chiral susceptibility in a finite volume.

\subsection{Phase structure in the current quark mass window}

As noted in the previous section we find in the entire range of
quark masses analyzed by us that the peak height of the chiral susceptibility
decreases with increasing volume. 
There is no hint for an increase of the peak height with volume, which is
what one would expect to happen in the vicinity of a second or first order 
phase transition.
This indicates that there is no first order phase transition in the system 
with values of the quark mass down to $m_q=0.0009375$.

\begin{figure}[htp]
\begin{center}
\includegraphics[width=.4\textwidth]{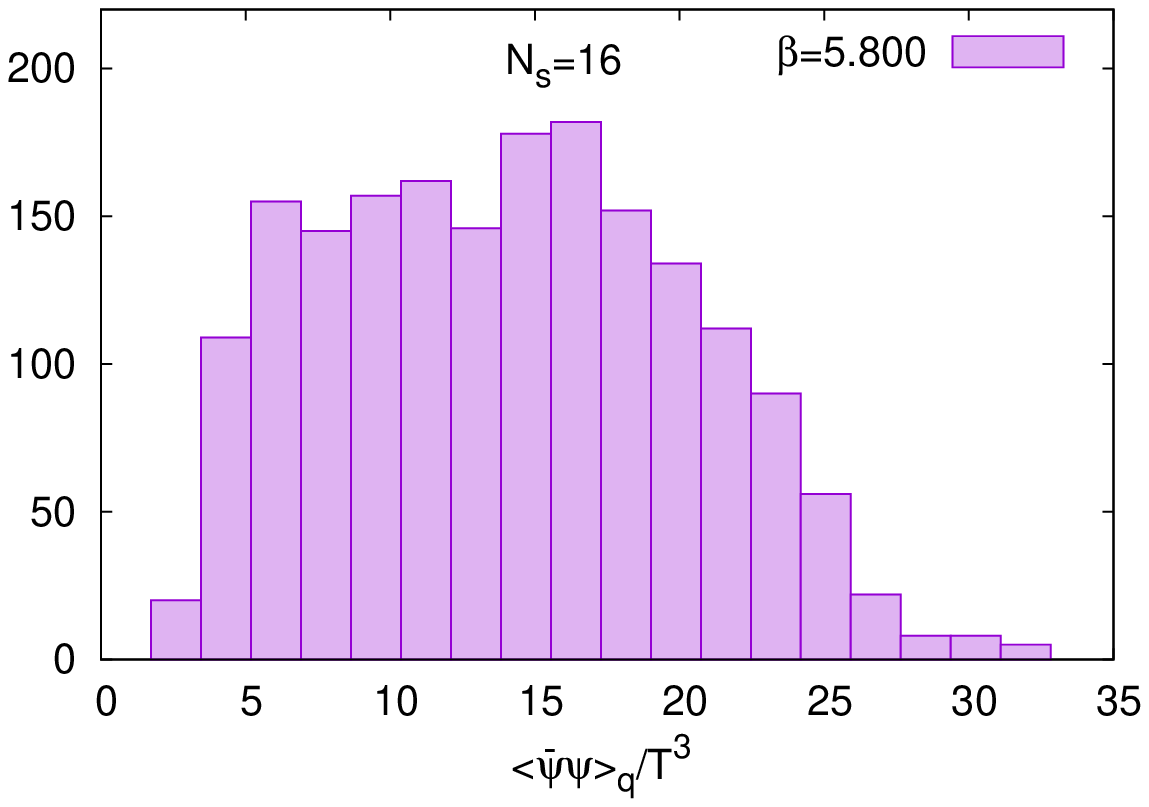}~~~\includegraphics[width=.4\textwidth]{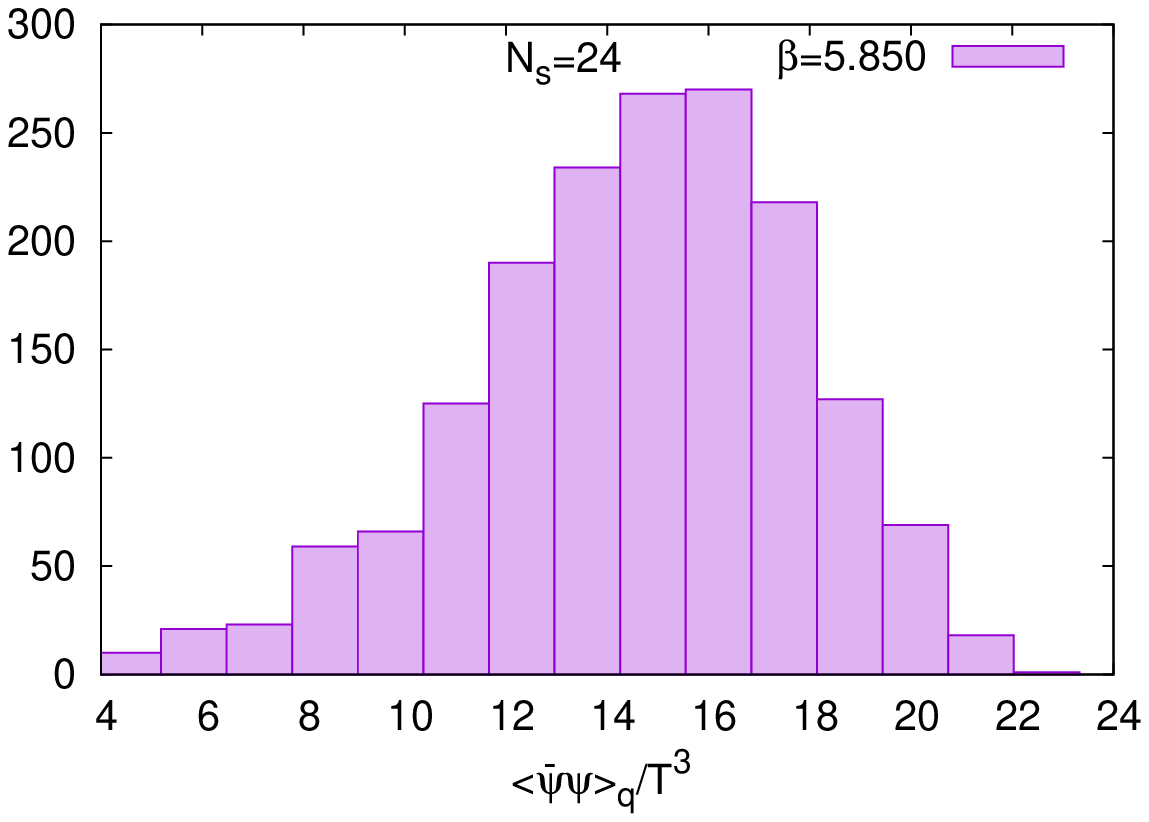}
\end{center}
\caption{The histogram of chiral condensates near $\beta_c$ with $m_q=0.0009375$ at $\beta=5.800$ on $16^3\times6$ (left) and at $\beta=5.850$ on $24^3\times6$ (right) lattices  .}
\label{fig:time_history}
\end{figure}

This is also supported by an analysis of the volume dependence of
histograms for the chiral condensate.
In Fig.~\ref{fig:time_history} such histograms are shown for the chiral
condensate at our lightest quark mass $m_q=0.0009375$ calculated for two different volumes, i.e. $N_s=16$
at $\beta=5.8$ and $N_s=24$ at $\beta=5.85$, which are close to the corresponding pseudo-critical
couplings for this value of the quark mass, $\beta_c(m_q)\simeq 5.78$ 
for $N_s=16$ and $\beta_c(m_q)\simeq 5.86$ for $N_s=24$.
There is no evidence that a double peak structure, which would be indicative
for the appearance of a first order phase transition, would develop in these
distributions as the volume increases. Thus there is no evidence for
two co-existing phases.

\begin{figure}[htp]
\begin{center}
\includegraphics[width=.45\textwidth]{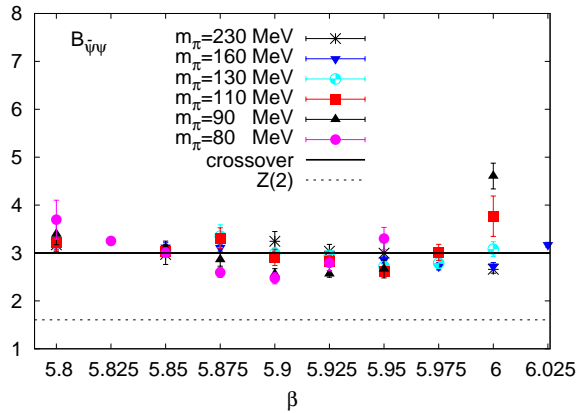}
\end{center}
\caption{The Binder cumulant of chiral condensates obtained from datasets with various values of pion masses.}
\label{fig:binder}
\end{figure}

We also analyzed the Binder cumulant $B_{\bar{\psi}\psi}$ of the chiral 
condensate at all values of the quark masses. $B_{\bar{\psi}\psi}$ is defined 
as follows
\be
B_{\bar{\psi}\psi} = \frac{\langle (\delta \bar{\psi}\psi)^4 \rangle} {\langle(\delta\bar{\psi}\psi)^2 \rangle^2}\; ,
\ee
where $\delta\bar{\psi}\psi=\bar{\psi}\psi-\langle\bar{\psi}\psi\rangle$
gives the deviation of the chiral condensate from its mean value
on a given gauge field configuration.
From different distributions of $\bar{\psi}\psi$ in different phases the value 
of $B_{\bar{\psi}\psi}$ can be obtained and can be used to distinguish phase 
transitions.
For a first order phase transition, $B_{\bar{\psi}\psi}=1$; for a crossover $B_{\bar{\psi}\psi}\simeq3$; for a second order transition belonging to the 3-d Z(2)
universality class, $B_{\bar{\psi}\psi}\simeq1.6$. As seen from Fig.~\ref{fig:binder}, the values of $B_{\bar{\psi}\psi}$ 
obtained from chiral condensates with different quark masses all lie around 3. There is a tendency for the lowest quark mass to give values smaller 
than 3 close to the crossover region. However, this clearly is not 
conclusive. Thus also the analysis of Binder cumulants presented
in Fig.~\ref{fig:binder} suggests that 3-flavor QCD with pion masses ranging 
from $230$ MeV to $80$ MeV corresponds to systems with
a smooth crossover transition.

In summary, all observables discussed above show no evidence for a 
first order phase transition in 3-flavor QCD with quark masses $m_q$ ranging 
from 0.0075 down to 0.0009375.

\subsection{Estimate of the critical pion mass}
\label{sec:estimate}

As discussed in the previous subsection we conclude that there is no first 
order phase transition even for quark masses as small as  
$m_q=0.0009375$ ($m_{\pi}\approx$ 80 MeV). 
However, we may test whether the chiral observables for different quark masses follow some specific 
scaling behaviors arising from the proximity of a chiral critical point. If such specific scalings are found to hold 
 for a window of quark masses then one can estimate bounds on the critical value for the pseudo-scalar 
 Goldstone mass. 
As we do not know the value of the critical quark mass in 3-flavor QCD,
it is at present not possible to determine the proper order parameter, e.g. from linear combinations of the chiral condensate and the gauge action, 
$M=\bar{\psi}\psi + s S_G$ \cite{Karsch:2001nf}. At present we thus
cannot perform a scaling analysis based on the construction
of the magnetic equation 
of state, c.f. Eq.~(\ref{eq:scaling}).
Instead, we will try to make use of the scaling behavior 
of the disconnected part of the chiral susceptibility at the pseudo-critical 
gauge coupling (temperature), i.e. we will use Eq.~(\ref{eq:chi_singular}) 
to estimate the critical pion mass.

First of all we need to determine the values of chiral observables at the 
pseudo-critical values, $\beta_c(m_q)$,  of the gauge coupling where 
the disconnected susceptibility peaks. We performed cubic spline fits to 
$\chi_{q,disc}/T^2$. 
Since we generally have about 8 data points ($\beta$ values) at each quark mass we mostly use 
three knots in the spline fit to keep a larger number of degrees of freedom, 
i.e. larger or equal to 3. Two of the knots
are fixed at the boundary of the $\beta$-range and the third knot is varied 
within the $\beta$-range. The fit results for chiral condensates and 
susceptibilities we have shown in the figures of this paper, e.g. in
Fig.~\ref{fig:obs_beta}, have been obtained by choosing the third knot such 
that the resulting $\chi^{2}/d.o.f$ is closest to unity. 
To determine the pseudo-critical couplings, $\beta_c$, 
we performed cubic spline fits for the disconnected part of the chiral 
susceptibilities using the bootstrap method,
where fits with $\chi^{2}/d.o.f$ closest to unity are chosen.
The value and error of the disconnected part of the chiral 
susceptibility at $\beta_c$ is obtained in the same way. 
The value $\beta_c$ for different quark masses and lattice sizes, the peak 
height of the disconnected part of the chiral susceptibility 
$\chi_{q,disc}/T^2$ and the value of the chiral condensate at $\beta_c$, 
$\langle \bar{\psi}\psi \rangle_q^c/T^3$, are listed in 
Table~\ref{tab:spline_fits}.  We also estimated systematic uncertainties of 
our fits by performing cubic spline fits of the disconnected part of the chiral 
susceptibility with the third knot chosen such that the $\chi^{2}/d.o.f$ 
is farthest away from unity. This results in values for
$\beta_c$ and $\chi_{\rm q,disc}^{c}/T^2$ that are within the uncertainties 
of those listed in Table~\ref{tab:spline_fits}. Note that the current way of performing spline fits also brings
in some artifacts, e.g. the unphysical dips seen in the right plots of Fig.~\ref{fig:obs_beta} and Fig.~\ref{fig:volume_dep9375} near the smallest $\beta$-values.
These dips can be cured by relocating the knots. However, the resulting changes of $\beta_c(m_q)$, $\langle \bar{\psi}\psi \rangle_q^c$ and $\chi^c_{q,disc}$ are small and well included within the uncertainties mentioned before. 

\begin{table}[t]
\small
\begin{center}
\begin{tabular}{|c|c|c|c|c|}
\hline
lattice dim. &   $m_q$     & $\beta_c(m_q)$      & $\chi_{\rm q,disc}^{c}/T^2$  &$ \langle \bar{\psi}\psi \rangle_q^c/T^3$\\
\hline
$16^3\times$ 6           & 0.0075          & 6.00(3)                       &  92(4)    &   15(2)               \\
$24^3\times$ 6          & 0.00375        & 5.941(8)                    &  155(6)        &     14.3(7)         \\
$16^3\times$ 6          & 0.00375        & 5.88(2)                    &  177(8)       &   18(2)                 \\
$12^3\times$ 6          & 0.00375        & 5.77(2)                    &  220(11)     &      24(2)              \\
$10^3\times$ 6          & 0.00375        & 5.70(2)                    &  246(10)     &     27(1)                \\
$24^3\times$ 6          & 0.0025        &  5.92(2)                     &   241(11)      &   14(2)                   \\
$24^3\times$ 6           & 0.001875      & 5.90(1)                   &  309(14)     &        13(1)         \\
$24^3\times$ 6          & 0.00125        &  5.87(1)                      &   463(21) &         14(1)               \\
$24^3\times$ 6           & 0.0009375    & 5.854(7)                     &   649(28)   &     14.0(7)                    \\
$16^3\times$ 6           & 0.0009375    & 5.77(1)                     &   737(35)   &            17(1)             \\
\hline
\end{tabular}
\end{center}
\caption{ A list of the pseudo-critical gauge coupling $\beta_c(m_q)$ and the peak height of the disconnected chiral susceptibility  $\chi^c_{q,disc}/T^2$
as well as the chiral condensate at the pseudo-critical gauge coupling $\langle \bar{\psi}\psi \rangle_q^c/T^3$ for various quark masses and volumes.
}
\label{tab:spline_fits}
\end{table}
\begin{figure}[htp]
\begin{center}
\includegraphics[width=0.46\textwidth]{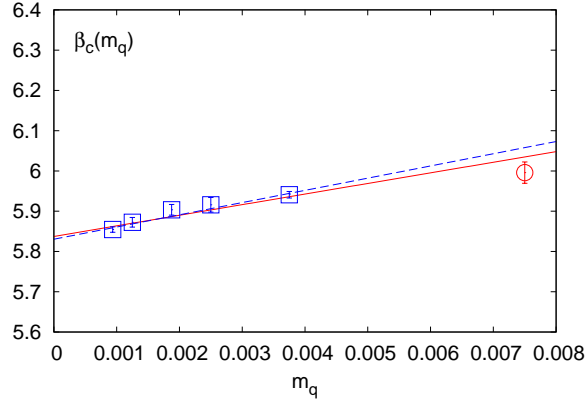}
\end{center}
\caption{$\beta_c$ as a function of quark mass $m_q$. The red solid and blue dashed lines represent linear fits using an ansatz of $-m_q/B+c$ with and without the data point at the heaviest quark mass, respectively.}
\label{fig:betac}
\end{figure}

Since in 3-flavor QCD the scaling variables are mixtures of $\Delta\beta$ and 
$\Delta m$ (c.f. Eq.~(\ref{eq:mixing})) also the scaling behavior of the order 
parameter in the quark mass
becomes complicated due to this mixture. In the immediate vicinity of the critical point, the t-direction
i.e. the line $h=0$ is defined as the tangent to the
pseudo-critical line~\cite{Karsch:2001nf}. In the limit $\Delta \beta
\rightarrow 0$ the mixing coefficient B can then be determined from

\begin{equation}
\frac{d \beta_c(m_q)}{d m_q} = - 1/B.
\label{eq:mixing_B}
\end{equation}

\begin{figure}[ht]
\begin{center}
\includegraphics[width=.48\textwidth]{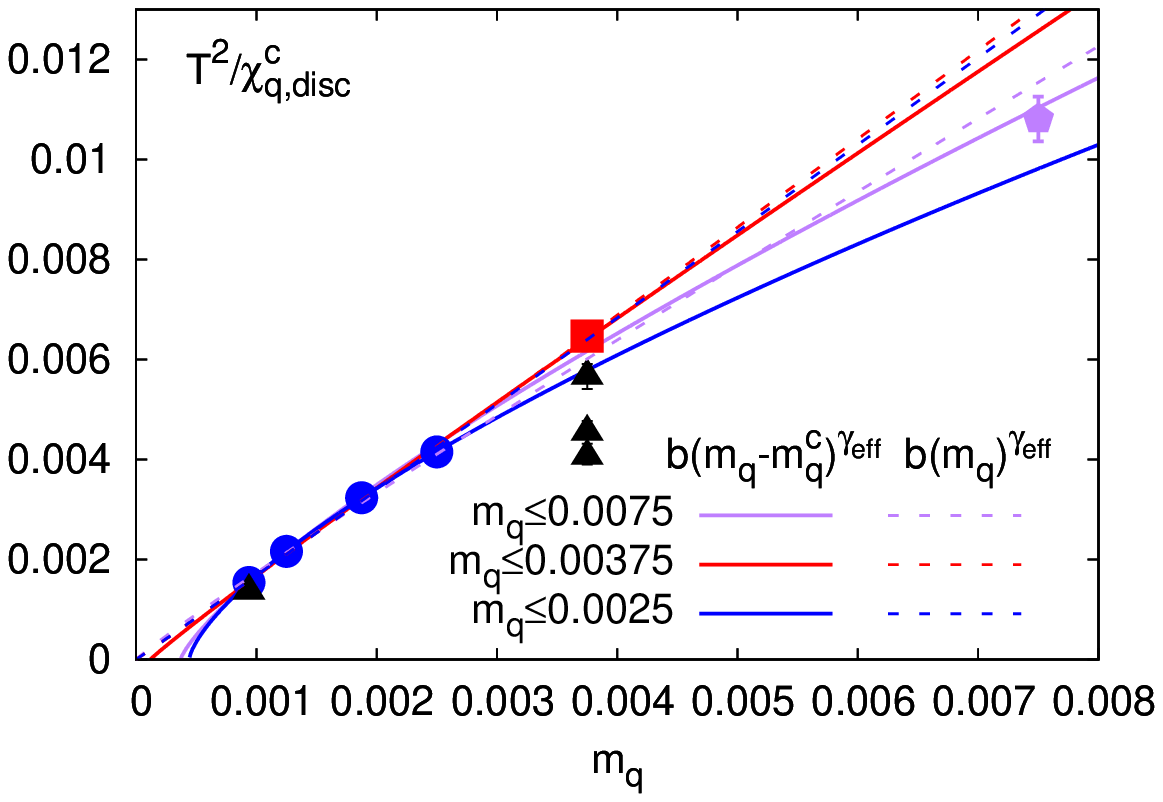}\includegraphics[width=.48\textwidth]{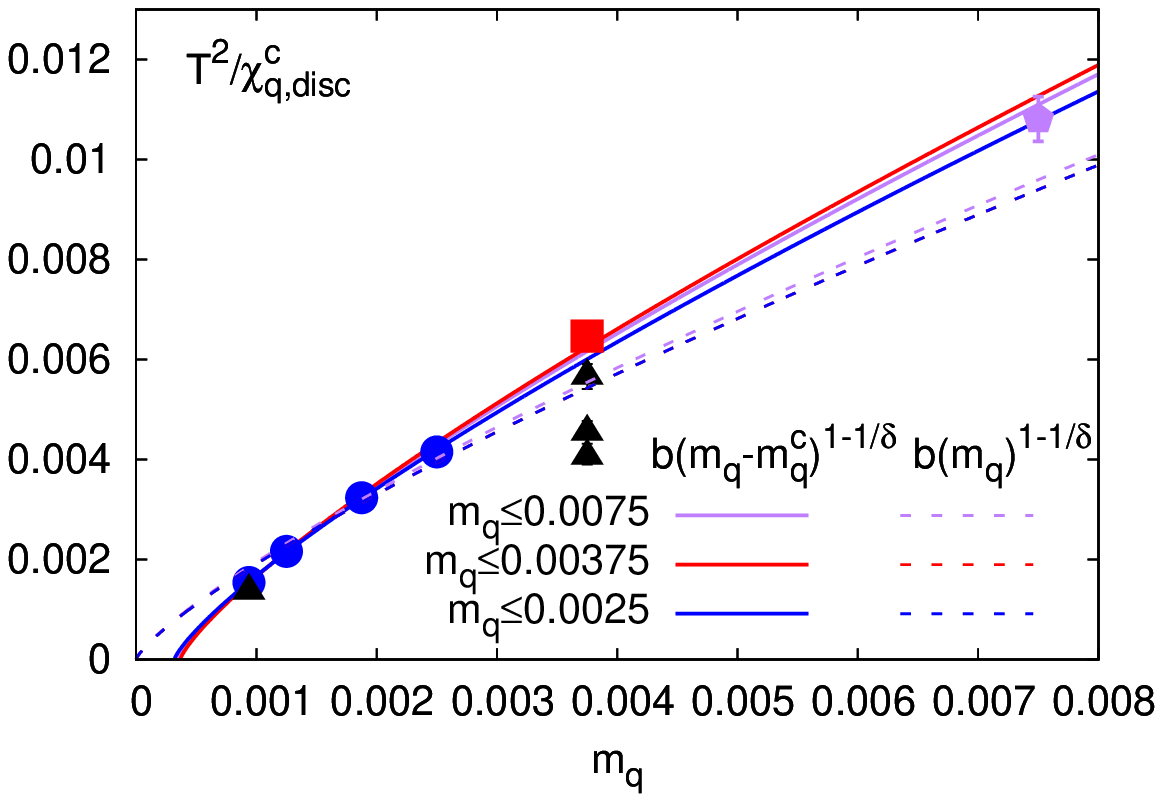}

\end{center}
\caption{The inverse of the maxima of the disconnected chiral 
susceptibility versus the bare quark mass, $m_q$. The solid lines and dashed lines show 
fits based on a scaling ansatz $b(m_q-m_q^c)^{exponent}$ and $b(m_q)^{exponent}$ , respectively.
The fit results shown in the left plot were obtained using an exponent of ${\gamma}_{eff}$ as a free parameter while in the right plot the exponent being fixed to ${1-1/\delta}$ . The fits with different
upper limits for the fit range in the quark mass, i.e. 0.0075, 0.00375 and 0.0025 are also shown.
The black triangles represent the data points for smaller volumes, i.e. $N_s$=16 with $m_q=0.0009375$ and $N_s=16,$ 12 and 10 with $m_q=0.00375$. 
For more details see discussions given in the text.}
\label{fig:mc}
\end{figure}

The dependence of the pseudo-critical coupling,
$\beta_c(m_q)$, on the quark mass is shown in Fig.~\ref{fig:betac}.
Shown also in Fig.~\ref{fig:betac} are two fits using as ansatz the leading linear term in Eq.~(\ref{eq:mb}) in 
two different fitting ranges.
These fits obviously yield an upper bound on
the absolute value of the mixing parameter $B$. 
From these fits we find that
$B$ is in the range of $-0.038\lesssim B\le 0$. Thus the scaling behavior of the chiral 
observables will be described by corresponding relations in Eqs.~(\ref{eq:chi_singular}) and 
(\ref{eq:M_singular}). However, the mixing is small. This has consequences for the
scaling of e.g. the peak in the chiral susceptibility. Although for 
any $B<0$ the peak height will diverge with the critical exponent $\gamma$, 
this behavior 
sets in only for very small values of the quark mass. In general one will
see an effective exponent $\gamma_{eff}$, which will closely resemble
the situation at $B=0$, i.e. $\gamma_{eff}\simeq 1 -1/\delta$. 
This can be seen from the fits to the peak heights of disconnected chiral susceptibilities with an ansatz of $b(m_q-m_q^c)^{\gamma_{eff}}$ where
$\gamma_{eff}$ is a fit parameter. The fit results are shown in the left plot of Fig.~\ref{fig:mc}. The fit to the
entire quark mass region is denoted by the purple solid line and it gives $\chi^2/d.o.f=0.88$ and ($m_q^c,\gamma_{eff}$)=(0.0004(1), 0.78(6)).
We also investigate the dependence of fit results on the fit range.  A fit leaving out the largest quark mass (denoted by the red solid line) yields 
$\chi^2/d.o.f=0.44$ and ($m_q^c,\gamma_{eff}$)=(0.00012(19)), 0.95(11)) while the fit leaving out the two largest quark masses (denoted by the blue solid line)
gives $\chi^2/d.o.f=0.004$ and ($m_q^c,\gamma_{eff}$)=(0.0004(2), 0.70(1)).

We thus have fixed $\gamma_{eff}$ to $1/\delta-1$ in the following analysis. In the right plot of Fig.~\ref{fig:mc} we show the fit results for $T^2/\chi_{q,disc}$
by using the ansatz $b\,(m_q-m_q^c)^{-1/\delta+1}$.
The critical quark mass $m_q^c$ can easily be obtained from the intercept of 
the fitting function 
and the quark mass axis. The fit to the whole quark mass region has a $\chi^2/d.o.f.$=0.67. It is 
shown in the right plot of Fig.~\ref{fig:mc} labeled by the purple solid line. The estimated critical quark mass
is $m_q^c=0.00035(3)$ with only the uncertainty from the fit.  
This corresponds to $m_\pi^c\simeq 50$ MeV. 
We also consider uncertainties arising from the fit range, i.e. 
the validity range of the scaling ansatz used for our fit. 
We did this by fitting to data without the data point
at the largest quark mass, i.e. $m_q=0.0075$.  The
$\chi^2/d.o.f.$ which resulted from the fit (denoted as the red solid line) without the data point at 
$m_q=0.0075$ remains almost the same. It gives a similar critical quark mass value
$m_q^c=0.00037(4)$.  While omitting the third heaviest quark mass in the fit results in a very small 
$\chi^2/d.o.f.$ i.e. 0.07. Nevertheless the obtained $m_q^c=0.00032(2)$ is consistent with previous two estimates within errors.
We also tried to fit the data with an ansatz motivated by the scaling behavior 
in the 3-dimensional Z(2) universality class but with a vanishing critical quark mass $m_q^c$. The ansatz thus is
$T^2/\chi_{q,disc}^c=b~m_q^{1-1/\delta}$ with only one fit parameter $b$.
Such a fit obviously cannot
describe the data at all~{\footnote{In the case $m_q^c$=0 one would, of course, expect that $O(N)$ critical exponents are more relevant than the Z(2). These critical exponents, however, are quite similar and would not change the conclusion given here.}, as can be seen from dashed lines in
the right plot of Fig.~\ref{fig:mc}. 
However, an ansatz of $bm_q^{\gamma_{eff}}$ can describe the data well when the largest two quark masses are excluded from the fit as shown in the left plot of Fig.~\ref{fig:mc}.
We thus cannot rule out that $m_q^c$ can actually be zero.

In Fig.~\ref{fig:mc} we also show $T^2/\chi_{q,disc}^c$ 
obtained for different volumes at the second highest and the smallest quark masses as the black triangles.
Since $T^2/\chi_{q,disc}^{c}$ reduces with decreasing volume, 
it is expected that effects arising from a finite volume overestimate $m_q^c$, i.e. in the thermodynamic limit 
$m_q^c$ becomes smaller.  Note that in our current simulations with $N_s=24$ even for the lightest quark mass
the finite volume effects are expected to be less than 10\% on $T^2/\chi_{q,disc}^c$ and consequently the change in the
estimate of $m_q^c$ in the thermodynamic limit is moderate.
With the limitations of finite volume effects and the fit results shown in Fig.~\ref{fig:mc}, at present, 
we can only provide an estimate for the upper bound of the critical pion mass, $m_\pi\simeq 50$ MeV~\footnote{We will nevertheless show in Appendix $B$ the fit results with a critical exponent of
$\gamma$ which supports current upper bound of $m_q^c$.}.
This result is compatible 
with that obtained from calculations using the stout action on $N_t=6$ lattices~\cite{Endrodi:2007gc}.

\begin{figure}[htp]
\begin{center}
\includegraphics[width=.55\textwidth]{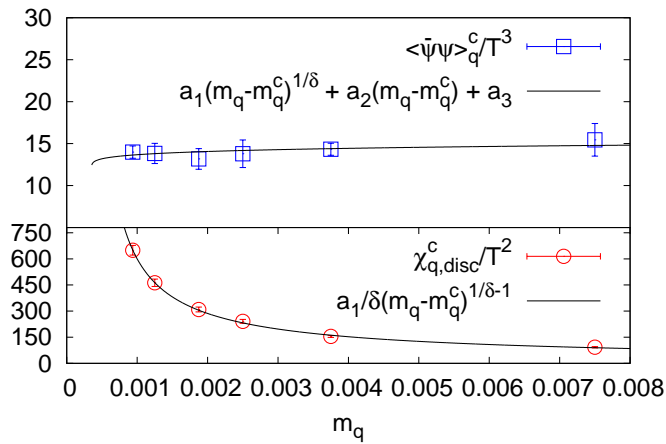}
\end{center}
\caption{A simultaneous joint fit using a 4-parameter ansatz to both light quark chiral condensate and the disconnected part of its susceptibility at $\beta_c$. 
 }
\label{fig:mc3}
\end{figure}

To have a better understanding
of the uncertainties in the estimate of $m_q^c$ we also take into account
possible contributions arising from the regular part of chiral observables, 
i.e. for the chiral condensate we use a fit ansatz of the form 
$a_1(m_q-m_q^c)^{1/\delta}+a_2 (m_q-m_q^c)+a_3$. 
Here, $a_2$ includes contributions from an $m_q$-type additive ultraviolet divergent term, as well 
as the regular part of the chiral susceptibility, and a non-vanishing $a_3$ reflects that the chiral order parameter
is not the correct order parameter and will approach a non-vanishing value
at $(m_q^c,\beta_c)$ . Since this involves four 
fit parameters and we only have six data points corresponding to the six 
different quark masses, we performed a joint fit to both the chiral 
condensate and the 
disconnected part of the chiral susceptibility. For the fit to the disconnected part of the chiral
susceptibility we assume that the singular behavior of chiral condensate is completely encoded within 
the disconnected chiral susceptibility, and not partly within the connected part of the chiral susceptibility. 
Further, the disconnected chiral susceptibility does not contain $1/a^2$ power-law divergences, and we assume that 
any additional contribution from the regular part of the chiral condensate to a diverging chiral susceptibility can be neglected. 
Thus, for the disconnected chiral susceptibility we use as a fit ansatz $a_1/\delta \cdot(m_q-m_q^c)^{1/\delta-1}$.
Results from such a combined fit
are shown in Fig.~\ref{fig:mc3}. This yields $a_1=8.6(2)$, $m_q^c=0.00035(3)$ and $a_3$=11.8(4) with $\chi^2/d.o.f=0.46$. 
The value of $a_1$ is consistent with $\delta/b$ where $b=0.55(1)$ is obtained from the fit shown in the right plot of Fig.~\ref{fig:mc}. 
The current estimate of $m_q^c$ is within the upper bound we obtained before, and the nonzero value of $a_3$ is certainly consistent 
with a nonzero value of $m_q^c$. The consistency can be understood since the chiral condensate is not a true order parameter as discussed in the Appendix $A$.
This estimate on $m_q^c$ is consistent with what we obtained from the fit shown in Fig.~\ref{fig:mc}. 
We also performed similar joint fits by taking into account a regular contribution to the disconnected chiral susceptibility with an 
ansatz $a_1/\delta  (m_q -  m_q^c )^{1/\delta-1} + a_4$.  Within errors, the fitted values of $a_1$ and $m_c$ were found to be consistent with the previous case while $a_4$ turned out to be vanishingly small compared to $a_2$.

\section{Conclusion}
\label{sec:summary}

We carried out calculations for 3-flavor QCD  
using the HISQ action on $N_{t}=6$ lattices. We used six values of pion masses in 
the mass range  $80~{\rm MeV} \lesssim  m_\pi \lesssim~$230 MeV.
 From the study of chiral condensates and chiral susceptibilities we found no 
direct evidence for the existence of a first order chiral phase transition in 
this pion mass region. Assuming that the quark masses used in this study lie within the critical scaling window of the anticipated 
chiral critical point of the 3-flavor QCD, we investigated 3-d Z(2) scaling behaviors of the chiral observables. 
Relying on these scaling studies, we were able to estimate an upper bound of the critical pion mass, i.e. $m_\pi\lesssim 50$ MeV.
As pointed out before, estimates of critical pion masses tend to yield smaller values as one approaches closer to the continuum limit, 
 either by going to finer lattice spacings or through using improved actions. While, in future, it will be essential to carry out lattice QCD calculations 
 for smaller quark masses and closer to the continuum limit to establish the first order chiral phase transition region of 3-flavor QCD, and it is likely 
 that this region will remain bounded by the critical pion mass estimated in the present study.  The estimated smallness of the critical pion mass 
 for 3-flavor QCD suggests that the first order chiral phase transition region of 3-flavor QCD might have little influence on the phase structure of 
 physical QCD, both for zero and non-zero baryon chemical potential.

\section*{Acknowledgments}
\label{ackn}

The numerical simulations were carried out on clusters of
the USQCD Collaboration in Jefferson Lab and Fermilab, and on BlueGene/L computers at the New York Center for Computational Sciences (NYCCS)
at Brookhaven National Lab, and on the local computing cluster at Central China Normal University. The work is partly supported 
by the U.S. Department of Energy under Contract No. DE-SC0012704, the Bundesministerium f\"ur Bildung und Forschung (BMBF)
under grant no. 05P15PBCAA and National Natural Science Foundation of China under grant numbers 11535012 and 11521064.

\section*{Appendix A: General behavior of order parameter and chiral condensate}

We want to discuss here the scaling behavior of the chiral condensate and
its susceptibility at a possible critical point in the coupling - quark mass 
plane which belongs to the 3-dimensional Z(2) universality class. At this critical point
the chiral condensate itself is not a true order parameter, but is part of
a mixture of operators that define the true order parameter $M$.

The behavior of the order parameter $M$ for this phase transition as a 
function of the quark mass at different values of temperature
is depicted in the left plot of Fig.~\ref{fig:order-parameter}.  
At fixed temperature the order parameter $M$ decreases 
when decreasing the external field $h$. From Eq.~(\ref{eq:mixing})
it follows that this corresponds to a decreasing quark mass,
\begin{equation}
m_q= h_0 h +m_q^c -B(\beta-\beta_c)\; .
\label{eq:m-h}
\end{equation}
The $h=0$ line is indicated in Fig.~\ref{fig:order-parameter} by a
dashed (red) line. For $B<0$ the lines of constant temperature, i.e. constant $\beta$, will
end for $h=0$ at coordinates of $(M>0,m_q<m_q^c)$ in the $M-m_q$ plane in the symmetry broken phase where $\beta<\beta_c$
and at $(M=0,m_q\ge m_q^c)$ in the symmetry restored phase where $\beta>\beta_c$. In particular at $T=T_c(m_q=m_q^c)$ and $m_q=m_q^c$
the line of constant temperature ends at $M=0$ marked by a big blue dot in the left plot of Fig.~\ref{fig:order-parameter}.
 When changing the external field $h$ from positive to 
negative values at $T<T_c(m_q=m_q^c)$ the order parameter $M$ will change discontinuously,
i.e. this temperature range corresponds to the first order transition region 
in 3-flavor QCD.  
Finally at temperatures lower than the critical temperature in the chiral limit, i.e. when $T<T_c(m_q=0)$
no phase transition occurs, irrespective of the value of quark mass.
For any value of the quark mass the system is in the spontaneously broken 
phase.

\begin{figure}[htp]
\begin{center}
\includegraphics[width=.5\textwidth]{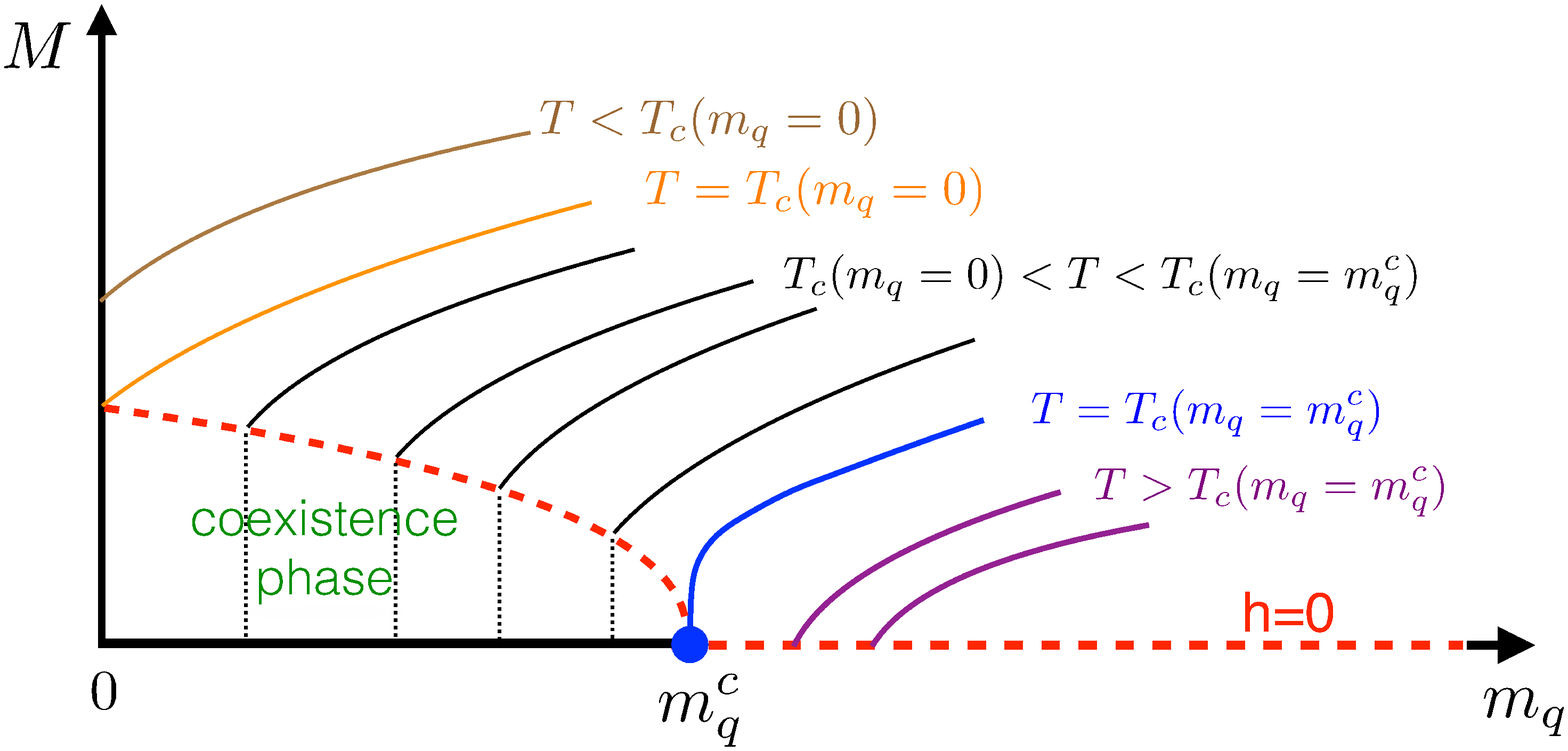}~~~\includegraphics[width=.45\textwidth]{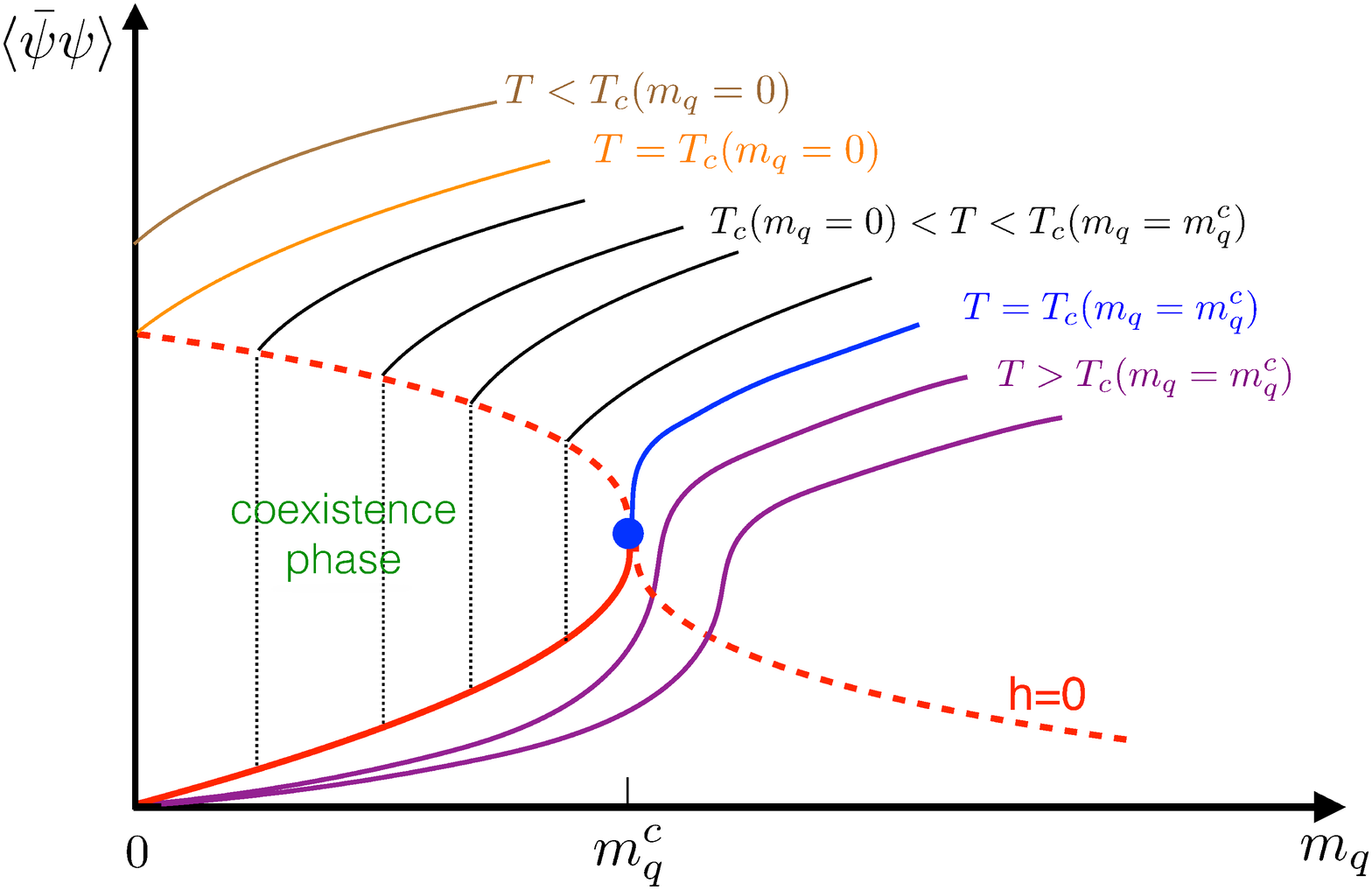}
\end{center}
\caption{Left: The order parameter $M$ of the phase transition in the 3-flavor QCD as a function of quark masses at different temperatures. Right: Same as the left plot
but for the chiral condensate. Here $T_c(m_q)$ is the (pseudo-) critical temperature of QCD with different values of quark masses $m_q$. 
The red dashed line indicates
the line of vanishing external field, $h=0$, and the blue dots mark the 
location of the critical point at $T=T_c(m_q^c)$.
}
\label{fig:order-parameter}
\end{figure}

The situation is similar for the quark chiral condensate 
$\langle\bar{\psi} \psi \rangle$, which obviously is not the order parameter 
for 3-flavor QCD. However, for small values of the mixing parameter $B$, 
it is a dominant component of the order parameter.  We sketch its behavior
in Fig.~\ref{fig:order-parameter}~(right). 
Of course, the relation between $m$ and $h$ given by Eq.~(\ref{eq:m-h}) remains
the same in this case. However, the y-coordinate for the line corresponding
to $h=0$ gets distorted because
$\langle\bar{\psi} \psi \rangle$ stays finite at  
the critical point $(\beta_c,m_c)$, which is marked also by a big blue dot 
in the right hand part of Fig.~\ref{fig:order-parameter}.
From this it is apparent that also for $T>T_c(m_q=m_q^c)$ and $m_q>m_q^c$ 
the chiral condensate $\langle\bar{\psi} \psi \rangle$ does not vanish
at $h=0$ as it is not the proper order parameter. The h=0 line is also indicated by a dashed line in the figure. Also note
that at the temperature corresponding to the chiral limit ($T=T_c(m_q=0)$)
the chiral condensate drops from a finite value directly to zero.

In general we have no direct access to the order parameter itself,
which needs to be constructed from e.g. a linear combination of
the chiral condensate and the gauge action. For our purpose,
however, it is sufficient to relate the chiral condensate and its 
susceptibility to the singular part of the free energy introduced
in Eqs.~(\ref{free}) and (\ref{sing}). For the contribution of the
singular part of the free energy to the chiral condensate we 
then obtain
\begin{equation}
\langle \bar{\psi}\psi\rangle_{sing} =  h^{1/\delta}
\tilde{f}_G(t,h)
\; , 
\end{equation} 
with $t$ and $h$ introduced in Eqs.~(\ref{eq:mixingt}) and (\ref{eq:mixing}), 
respectively.
For $A=B=0$ the function $\tilde{f}_G(t,h)$ reduces to the scaling function 
$f_G(z)$. However for $A\ne 0$, $t$ as well as $h$ depend on the quark
mass and $\tilde{f}_G$ thus receives contributions from
partial derivatives of both arguments of $f_{sing}$

\begin{eqnarray}
\tilde{f}_G(t,h) &=& 
-\Big( 1+\frac{1}{\delta}\Big) f_s(z) -\frac{1}{\beta\delta} 
\frac{\partial z}{\partial m_q}
\frac{\partial f_s(z)}{\partial z} \nonumber \\
&=& -\Big( 1+\frac{1}{\delta}\Big) f_s(z) - 
\frac{z}{\beta\delta} \Big( A \beta\delta \frac{h}{t} -1\Big)  
\frac{\partial f_s(z)}{\partial z} \nonumber \\
&=& f_G(z) - A h^\omega \frac{\partial f_s(z)}{\partial z} \frac{h_0}{t_0}\;\; ,\;\; 
\label{eq:psising}
\end{eqnarray}
with $\omega = 1-1/\beta\delta$. In general, the singular part of the free 
energy thus does not
lead to a simple scaling form of the chiral condensate. It receives 
corrections to scaling, which relative to the $h$-dependence of the
chiral condensate, are suppressed by a factor $h^\omega\simeq h^{1/2}$.

Similarly we can analyze the scaling properties of the chiral 
susceptibility, 
\begin{equation}
\chi_q = \frac{\partial \langle \bar{\psi}\psi \rangle}{\partial m_q} \; .
\end{equation}
The singular contribution to $\chi_q$ can then be obtained from 
Eq.~(\ref{eq:psising}),
\begin{eqnarray}
\chi_q^{sing} &=& \frac{\partial \langle \bar{\psi}\psi \rangle_{sing}}{\partial m_q} 
= \frac{1}{h_0} h^{1/\delta-1} \tilde{f}_\chi(t,h) \; ,
\label{eq:chising}
\end{eqnarray}
with
\begin{equation}
\tilde{f}_\chi(t,h) = f_\chi(z) + A h^\omega \,\frac{h_0}{t_0}P_1(z) + \left(A h^\omega\,\frac{h_0}{t_0}\right)^2 P_2(z)
\; ,
\end{equation}
and
\begin{eqnarray}
P_1(z)  &=&  f'_G(z) -(\omega+\frac{1}{\delta})f'_s(z) + \frac{z}{\beta\delta} f''_s(z) \; ,
\nonumber \\
P_2(z)  &=& - f''_s(z) \; .
\end{eqnarray}
One thus finds that the chiral susceptibility diverges in the vicinity
of the critical point just like the order parameter susceptibility, 
i.e. $\chi_q \sim h^{1/\delta -1}$, as shown in Eq.~(\ref{eq:chi_singular}).

In order to make sure that the scaling arguments given in connection
with Eq.~(\ref{eq:chi_singular}) are valid also for a pseudo-critical
coupling extracted from the location of a peak in the chiral susceptibility
rather than the true order parameter $M$,
we also need to check that the locaton of this peak 
corresponds to a constant value of the scaling variable $z$. I.e. in the
chiral limit the peak is located at some position $z=z_c$.
Ignoring the corrections to scaling in Eq.~(\ref{eq:chising}) we may determine 
the location of the peak of $\chi_q$ at fixed quark mass. 
We need to solve
\begin{eqnarray}
0 &=& \frac{\partial}{\partial \beta } h^{1/\delta -1} f_\chi(z) \nonumber \\
&=& B \left( \frac{1}{\delta}-1\right) f_\chi(z) - 
\left( h^\omega - \frac{B}{\beta\delta} z\right)  f'_\chi (z)
\; ,
\label{eq:zero}
\end{eqnarray} 
Obviously for $B=0$ the peak of $\chi_q$ at fixed quark mass 
corresponds to the peak of $f_\chi(z)$, i.e. $z_p$. 
For $B\ne 0$ Eq.~(\ref{eq:zero}) is a function of the scaling variable
$z$ aside from some corrections to scaling that are proportional 
to $h^\omega$. The location of the peak in the chiral susceptibility thus  
is controlled by a value of the scaling variable $z=z_c$ up to some 
corrections to scaling that vanish in the limit $h\rightarrow 0$. For
small values of $B$ we can determine $z_c$ by expanding around $z=z_p$, i.e.
$\Delta z/z_p=z_c/z_p-1=B(1/\delta-1)f_\chi(z_p)\Big{/}\left(z_p^2 f_\chi^{\prime\prime}(z_p) (\frac{h}{t}|_{z_p}\frac{h_0}{t_0}-\frac{B}{\beta\delta})\right)$.

\section*{Appendix B: Mixing coefficient and effective critical exponent}

As seen from Eq.~(\ref{eq:mb}), the
dominant term determining the relation between $\Delta \beta$ and $\Delta m$
in the limit $\Delta \beta \rightarrow 0$ is proportional
to $B$, which will be small in QCD when the critical quark mass is small.
Outside a small asymptotic scaling region the second term in Eq.~(\ref{eq:mb})
thus will be dominant and $\chi_M$ may show an effective scaling controlled
by the exponent $1-1/\delta$. To illustrate this we consider the simple 
scaling form

\begin{equation}
\chi_M^{peak}  \sim h^{1/\delta-1} \sim  \left( (1+B^2) \Delta \beta - B
\left( z_0/z_p\, (1+B^2)  \Delta \beta \right)^{\beta\delta}\right )^{-\gamma}
\; ,
\label{eq:hansatz}
\end{equation}
where we used Eqs.~(\ref{eq:zp}) and~(\ref{eq:mb}) to express the external field $h$, introduced
in Eq.~(\ref{eq:mixing}), in terms of $\Delta \beta$ only.

\begin{figure}[htp]
\begin{center}
\includegraphics[width=.45\textwidth]{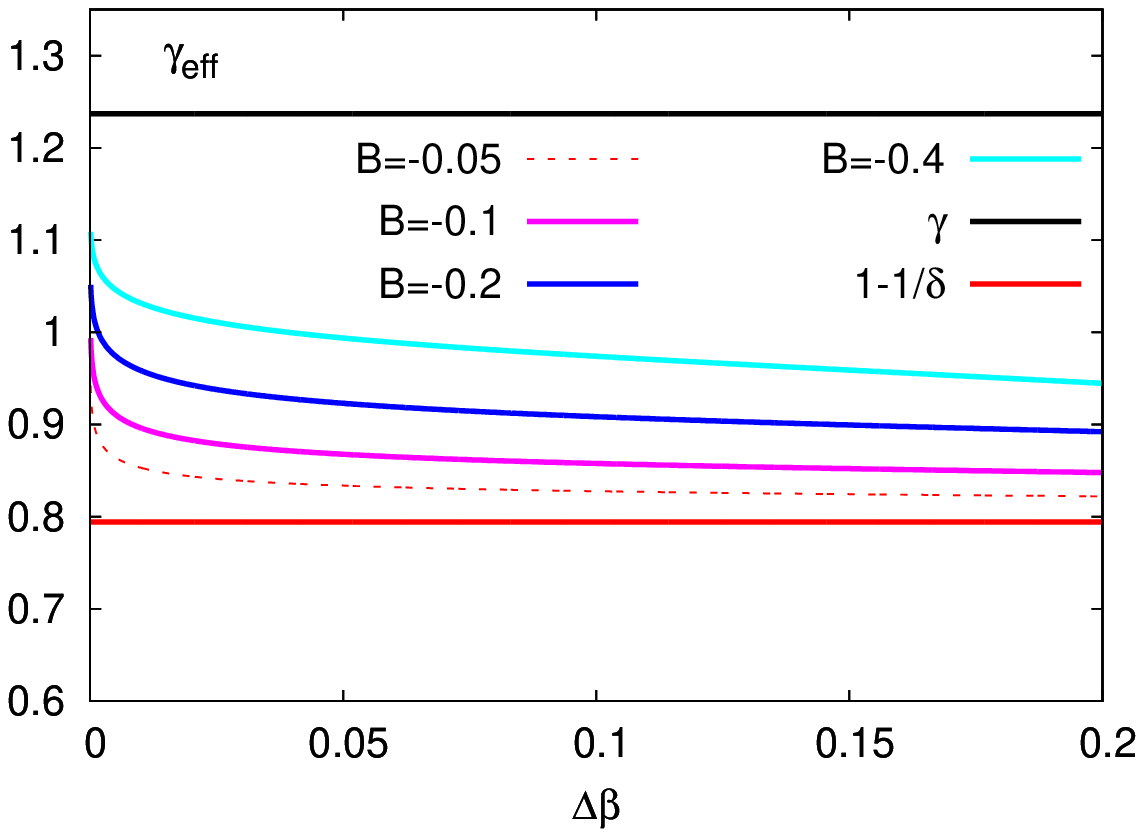}\includegraphics[width=.45\textwidth]{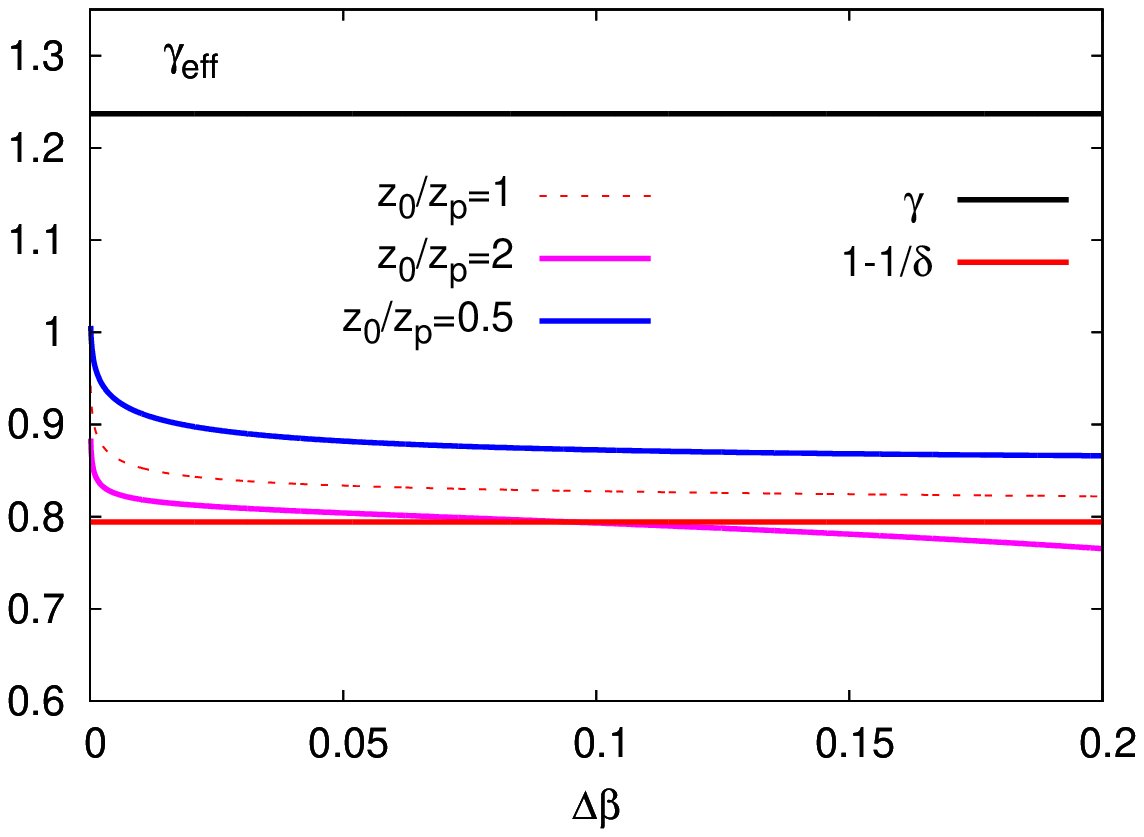}
\end{center}
\caption{The effective scaling exponent, 
$\gamma_{eff}= -\ln (\chi_M^{peak})/\ln(\Delta m)$, 
extracted from the peak value of the order parameter 
susceptibility obtained from Eq.~(\ref{eq:hansatz}). Here $\gamma_{eff}=\gamma$ when B=0. Left: For $z_0/z_p=1$ and several 
values of the mixing coefficient $B$. Right: For $B=-0.05$ and several different values of $z_0/z_p$.}
\label{fig:gammaeff}
\end{figure}

\begin{figure}[htp]
\begin{center}
\includegraphics[width=.45\textwidth]{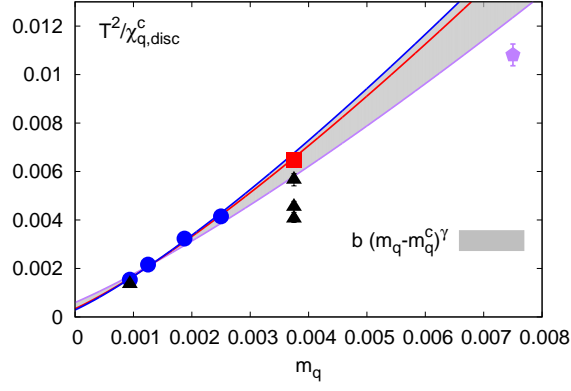}
\end{center}
\caption{Fits to $T^2/\chi^{c}_{q,disc}$ using an ansatz of $b(m_q-m_q^c)^{\gamma}$. The purple, red and blue line represents the fit to the whole quark mass region, $m_q<0.0075$ and $m_q<0.00375$, respectively.}
\label{fig:fit_gammaeff}
\end{figure}

We show in Fig.~\ref{fig:gammaeff} the effective exponent $\gamma_{eff}$
describing the behavior of $\chi_M^{peak}$ as function of $\Delta \beta$.
This shows that we may expect to find a rather complicated scaling behavior
of $\chi_M^{peak}$. We can see that as long as the value of B is small the effective
exponent is much closer to $1-1/\delta$ rather than $\gamma$ in most cases that might apply to our current
investigation .

As discussed above and also in Section~\ref{sec:estimate} the effective critical exponent is 
expected to be rather close to $1/\delta-1$. 
We nevertheless would like to check the uncertainties in the estimate 
of the critical quark mass arising from the critical exponents. 
We thus use an ansatz of $b(m_q-m_q^c)^{\gamma}$ to fit the inverse of the 
disconnected part of the chiral susceptibilities at $\beta_c(m_q)$.
The fit result is shown in Fig.~\ref{fig:fit_gammaeff}. All the fits to the data sets with all the masses, without the heaviest mass point and without two heaviest mass points prefer negative values of 
$m_q^c$ which can be seen clearly from the intercept of the solid lines 
with the $x$ axis.

\bibliographystyle{JHEP}

\providecommand{\href}[2]{#2}\begingroup\raggedright\endgroup
\end{document}